\documentclass[twocolumn,tighten,times,trackchanges]{aastex63} 
\usepackage{subfigure,graphicx}
\usepackage{url}
\usepackage{color}
\definecolor{darkgray}{gray}{0.4}
\definecolor{patinared}{rgb}{.72,0,0}
\definecolor{patinablue}{rgb}{0,0,.65}
\definecolor{orange}{rgb}{1,0.5,0}

\hypersetup{
  colorlinks= true, 
}
\usepackage{amsmath}
\usepackage{xfrac}

\renewcommand{\ion}[2]{#1$\,${\textsc{\romannumeral #2}}}
\newcommand{\lt}{\textsc{Little Things}}
\newcommand{\herschel}{\textit {Herschel}}
\newcommand{\spitzer}{\textit {Spitzer}}
\newcommand{\scanam}{\textsc{Scanamorphos}}
\newcommand{\hi}{\ion{\rm{H}}{1}}
\newcommand{\hii}{\ion{\rm{H}}{2}}
\newcommand{\htwo}{\rm{H}$_2$}
\newcommand{\cii}{[\ion{\rm{C}}{2}]}
\newcommand{\oi}{[\ion{\rm{O}}{1}]}
\newcommand{\um}{ $\mu$m}

\defcitealias{RR13}{RR+13}

\shorttitle{\herschel\ Photometry of \lt\ Dwarfs}   
\shortauthors{Cigan et al.} 

\submitjournal{\aj}

\begin{document}

\title{\herschel\ Photometric Observations of \lt\ Dwarf Galaxies}

\author[0000-0002-8736-2463]{Phil Cigan}
\affiliation{George Mason University, 4400 University Dr, Fairfax, VA 22030-4444, USA}

\author[0000-0002-5669-5038]{Lisa M. Young}
\affiliation{New Mexico Institute of Mining and Technology, 801 Leroy Place, Socorro, NM 87801, USA}

\author[0000-0003-3398-0052]{Haley L. Gomez}
\affiliation{School of Physics and Astronomy, Cardiff University, Cardiff CF24 3AA, UK}

\author{Suzanne C. Madden}
\affiliation{CEA Saclay, Laboratoire AIM, CEA/DSM-CNRS-Universit\'e Paris Diderot DAPNIA/Service d'Astrophysique, B\^at. 709, CEA-Saclay, 91191 Gif-sur-Yvette Cedex, France}

\author{Pieter De Vis}
\affiliation{School of Physics and Astronomy, Cardiff University, Cardiff CF24 3AA, UK}

\author[0000-0002-3322-9798]{Deidre A. Hunter}
\affiliation{Lowell Observatory, 1400 West Mars Hill Road, Flagstaff, AZ 86001, USA}

\author[0000-0002-1723-6330]{Bruce G. Elmegreen}
\affiliation{IBM T. J. Watson Research Center, P.O. Box 218, Yorktown Heights, New York, USA}

\author[0000-0002-7758-9699]{Elias Brinks}
\affiliation{Centre for Astrophysics Research, University of Hertfordshire, College Lane AL10 9AB, UK}

\collaboration{40}{and the \lt\ Team}

\correspondingauthor{Phil Cigan}
\email{pcigan@gmu.edu}

\begin{abstract}

We present here far-infrared photometry of galaxies in a sample that is relatively unexplored at these wavelengths: low-metallicity dwarf galaxies with moderate star formation rates.  
Four dwarf irregular galaxies from the \lt\ survey are considered, with deep \herschel\ PACS and SPIRE observations at 100\um, 160\um, 250\um, 350\um, and 500\um.  
Results from modified-blackbody fits indicate that these galaxies have low dust masses and cooler dust temperatures than more actively star-forming dwarfs, occupying the lowest $L_\mathrm{TIR}$ and $M_\mathrm{dust}$ regimes seen among these samples.  
Dust-to-gas mass ratios of $\sim$10$^{-5}$ are lower, overall, than in more massive and active galaxies, but are roughly consistent with the broken power law relation between the dust-to-gas ratio and metallicity found for other low-metallicity systems.  
Chemical evolution modeling suggests that these dwarf galaxies are likely forming very little dust via stars or grain growth, and have very high dust destruction rates.  

\end{abstract}

\section{INTRODUCTION}
\label{sec:intro}

Dust plays an important role in the process of star formation and, therefore, evolution of galaxies.
Combined with metallicity measurements, the dust content can provide a probe of the evolutionary status of a galaxy (e.g., \citealp{Clark2015,Schneider2016}). Historically, studies have suggested that the dust-to-gas ratio linearly increases with the content of metals \citep{Dwek1998,Edmunds2001,Draine2007} as a galaxy uses up its gas to form stars. Far infrared (FIR) to sub-millimeter (sub-mm) photometry can directly probe the emission from dust in dwarf galaxies to help answer a fundamental question: how do the bulk dust properties (mass, temperature, emissivity) vary with metallicity ($Z$) and what are the dust scaling relations for galaxies of different morphological type and in different evolutionary phases?   
Constraining dust properties requires robust sampling of the thermal emission profile -- with sensitive measurements in the infrared to sub-mm, extending beyond the peak to longer wavelengths -- as well as reliable dust and spectral energy distribution (SED) models.
IR wavelengths are also difficult to observe from ground-based instruments due to water vapor absorption features in the atmosphere, so they can only detect the brightest (highest metallicity or extreme star forming) dwarfs.  The \herschel\ {\it Space Observatory} (hereafter \herschel, \citealp{Pilbratt2010}) (and its sister satellite {\it Planck}, \citealp{planckI}) has made it possible to study the FIR--sub-mm properties of local galaxies in large samples (e.g., \citealp{Skibba2011,Boselli2010,Leroy2011,Dunne2011,Negrello2013,Clemens2013,Agius2013,RemyRuyer2014, RomanDuval2014,Clark2015,Beeston2018}) allowing for the first derivations of dust scaling relations and comparison of dust properties across the Hubble Sequence with coverage over the full FIR spectrum \citep{Cortese2012,Smith2012a,DeVis2017a}, as well as resolved dust studies of nearby galaxies \citep{Smith2012b,Galametz2011,Gordon2014,Draine2014}. Many of these studies sample the more FIR-bright spirals or more massive early type galaxies, but in the low stellar mass and low metallicity regime, dwarf galaxies can provide a fundamental probe of the dust content in sources that are chemically young and potentially representative of galaxies in the early universe.

The largest \herschel\ sample of dwarfs to date comes from the Dwarf Galaxy Sample (hereafter DGS) of \citet{DGS}, where \citet{RemyRuyer2014} indicated that the dust-to-gas ratio (DGR) is linearly proportional to metallicity from the solar level down to $Z\sim 0.2\,Z_{\odot}$, but the slope steepens for lower metallicities such that these galaxies are predicted to have less dust for a given gas mass compared to higher metallicity sources; at $Z=0.025\,Z_{\odot}$ the dust can be lower by an order of magnitude or more (see also \citealp{Draine2007,Galliano2008}). This result is also suggested by \citet{DeVis2017b}, who combined the DGS with galaxies from the \herschel-ATLAS survey \citep{Eales2010}, roughly doubling the number of low metallicity galaxies (though many were not detected in the sub-mm with \herschel).  Difficulty in determining the origin of the differences in dust mass and/or physical properties between low and high metallicity galaxies is further exacerbated due to the enormous scatter in the DGR at low $Z$, with \citet{Hunt2014} demonstrating that even with the same metallicity, the observed DGR in two low-$Z$ dwarf galaxies can vary by two orders of magnitude (see also \citealp{Schneider2016}).   
In \citet{RemyRuyer2014} and \citet{DeVis2017b}, chemical evolution models are used to probe the observed differences in dust content with metallicity and the deviation from the linear trend seen in the DGR for higher metallicity galaxies. Those works attribute the variation to different sources of dust (i.e., different amounts of star-dust from evolved stellar winds and/or supernovae, and grain growth in the interstellar medium), whereas it is instead attributed to different star formation rates in \citet{Zhukovska2014} and the balance between inflows and outflows in \citet{Feldmann2015}. \citet{Schneider2016} convincingly make the case that the higher dust masses observed in the low $Z$ dwarf SBS 0335-052, as compared to I Zw18 which has the same metallicity, is due to increased dust grain growth in the ISM in SBS 0335-052 because of higher interstellar gas densities.

Understanding what underpins the relationship between the dust content of galaxies and metallicity and gas properties is therefore important, and the low metallicity regime seems crucial for revealing trends that are different compared to normal spirals. 
In this paper, we provide a sample of deep FIR photometric measurements using \herschel\ for four low metallicity, low star forming galaxies from the \lt\ survey \citep{Hunter2012LTdata}. This provides additional sources to the dwarf studies in the literature, crucially sampling a regime that still lacks significant numbers: that of low mass, moderate star formation rate, and extremely low metal content.
We use this information to analyze the physical properties of the ISM in these galaxies, such as dust temperature, mass, emissivity, and infrared luminosity, and compare to other galaxy samples.  Information about the sample, the observations, and data reduction are discussed in \S\ref{sec:data}.  The photometry methods, including selection of regions and extraction of flux densities, are described in \S\ref{sec:photometry}.  In \S\ref{sec:Bandcomp}, we compare the \herschel\  observations with other IR data.  \S\ref{sec:SEDs} introduces the SED model used to derive the properties of these sources, with comparison to the literature discussed in \S\ref{sec:fitresults}.

\section{DATA}
\label{sec:data}

\subsection{The Sample}
\label{sec:sample}

Four dwarf irregular galaxies with properties typical of normal dwarfs were selected for observations with \herschel\ for this study: DDO 69, DDO 70, DDO 75, and DDO 210.  The photometry sample is slightly different from the spectroscopy sample of \cite{Cigan2016} --  DDO 69, DDO 70, and DDO 75 are common between the two, but DDO 210 has no \herschel\ emission line data.  They are all nearby (less than 1.3 Mpc) and small (R$_D$ between 0.17 and 0.48 kpc, $M_\mathrm{HI}$ of 7.2$\times 10^{7}$ $M_\odot$ or less), with modest star formation rates of 0.012 $M_\odot$ yr$^{-1}$ or less. The most metal rich galaxy in the sample is DDO 75, at 12+log(O/H) = 7.54 (7\% $Z_\odot$), while the poorest is DDO 210 at 7.2 (3.2\% $Z_\odot$).  A summary of the parameters for the sample is given in Table~\ref{table:sampledata}.

For comparison with other local dwarf galaxies, the DGS metallicities span a range of 7.14--8.43 \citep{DGS} and star formation rate (SFR) values range several orders of magnitude from 0.0008--43 $M_\odot$ yr$^{-1}$ \cite{DeLooze2014}.

\begin{deluxetable*}{ llccccccc }[t!]
\tabletypesize{\footnotesize}

\tablecaption{ Sample Galaxy Parameters \label{table:sampledata}} 

\tablehead{ \colhead{Galaxy} & \colhead{Other Names} & \colhead{D} & \colhead{log$_{10}$ $M_\mathrm{HI}$} & \colhead{R$_D$} & \colhead{${\mu_0}^V$} & \colhead{log$_{10}$ SFR$^{FUV}$} & \colhead{log$_{10}$ SFR$_D^{FUV}$} & \colhead{12+log$_{10}$(O/H)} \\
 & & (Mpc) & ($M_\odot$) & (kpc) & (mag arcsec$^{-2}$) & ($M_\sun \; \textrm{yr}^{-1}$) & ($M_\sun \; \textrm{yr}^{-1} \; \textrm{kpc}^{-2}$) & }

\startdata
DDO 69 & PGC 28868 & 0.8 & 6.84 & 0.19 $\pm$ 0.01 & 23.01 & -3.17 & -2.22 $\pm$ 0.01 & 7.38 $\pm$ 0.10 \\
       & UGC 5364 \\
       & Leo A    \\
DDO 70 & PGC 28913 & 1.3 & 7.61 & 0.48 $\pm$ 0.01 & 23.81 & -2.30 & -2.16 $\pm$ 0.00 & 7.53 $\pm$ 0.06 \\
       & UGC 5373 \\
       & Sextans B\\
DDO 75 & PGC 29653 & 1.3 & 7.86 & 0.22 $\pm$ 0.01 & 20.40 & -1.89 & -1.07 $\pm$ 0.01 & 7.54 $\pm$ 0.06 \\
       & UGCA 205 \\
       & Sextans A\\
DDO 210 & PGC 65367 & 0.9 & 6.3 & 0.17 $\pm$ 0.01 & 23.77 & -3.75 & -2.71 $\pm$ 0.06 & 7.2 $\pm$ 0.5 \\
        & Aquarius Dwarf  \\
\enddata

\tablerefs{Data as reported in \cite{Hunter2012LTdata}. Original distance and metallicity references, respectively. DDO 69: \cite{Dolphin2002}, \cite{VanZee2006}.  DDO 70: \cite{Sakai2004}, \cite{Kniazev2005}.  DDO 75: \cite{Dolphin2003}, \cite{Kniazev2005}.  DDO 210: \cite{Karachentsev2002}, \cite{Richer1995}. } 
\tablecomments{General information about this galaxy sample, as reported by \cite{Hunter2004Ha,Hunter2006UBV,Hunter2012LTdata}.  ${\mu_0}^V$ is the central $V-$band brightness.  R$_D$ is the disk scale length. SFR$^{FUV}$ is the star formation rate determined from $L_{FUV}$, and SFR$_D^{FUV}$ is that divided by $\pi$R$_D^2$.  Oxygen abundances for metallicities were determined from \hii\ regions. \\
}

\end{deluxetable*}

\subsection{Observations}
\label{sec:observations}

\begin{deluxetable*}{ llcccccc }[]
\tablecaption{Summary of \herschel\ Observations \label{table:Obs}}
\tablehead{ \colhead{Galaxy} & \colhead{Instrument} & \colhead{Filters} & \colhead{OBSID} & \colhead{RA (J2000)} & \colhead{DEC (J2000)} & \colhead{Duration} & \colhead{Map Size}  \\
			 & & ($\mu$m) & & (h m s) & (d m s) & (s) & (arcmin) }
\startdata
DDO 69  & SPIRE & 250, 350, 500 & 1342255183 & 09 59 29.30 & +30 44 21.17 & 2019 & 9.2 $\times$ 9.2 \\
        & PACS  & 100, 160      & 1342255335 & 09 59 26.44 & +30 44 43.37 & 4490 & 6.9 $\times$ 6.9 \\
        & PACS  & 100, 160      & 1342255336 & 09 59 26.73 & +30 44 47.40 & 4490 & 6.9 $\times$ 6.9 \\ \\
DDO 70  & SPIRE & 250, 350, 500 & 1342255156 & 10 00 01.09 & +05 19 57.42 & 2047 & 9.8 $\times$ 9.8 \\
        & PACS  & 100, 160      & 1342255962 & 10 00 00.09 & +05 19 56.02 & 4714 & 7.4 $\times$ 7.4 \\
        & PACS  & 100, 160      & 1342255963 & 10 00 00.11 & +05 19 56.27 & 4714 & 7.4 $\times$ 7.4 \\ \\
DDO 75  & SPIRE & 250, 350, 500 & 1342247243 & 10 10 59.88 & -04 41 31.73 & 2095 & 10.8 $\times$ 10.8 \\
        & PACS  & 100, 160      & 1342247428 & 10 11 00.79 & -04 41 34.31 & 6172 & 8.1 $\times$ 8.1   \\
        & PACS  & 100, 160      & 1342247429 & 10 11 00.78 & -04 41 34.26 & 6172 & 8.1 $\times$ 8.1 \\ \\
DDO 210 & SPIRE & 250, 350, 500 & 1342245438 & 20 46 51.48 & -12 51 15.73 & 1201 & 4.8 $\times$ 4.8 \\
        & PACS  & 100, 160      & 1342245178 & 20 46 51.81 & -12 50 52.50 & 2325 & 3.1 $\times$ 3.1 \\
        & PACS  & 110, 160      & 1342245179 & 20 46 51.80 & -12 50 51.78 & 2325 & 3.1 $\times$ 3.1 \\
\enddata

\end{deluxetable*}

\begin{figure*}[t!]
\centering
    \includegraphics[width=1.0\textwidth]{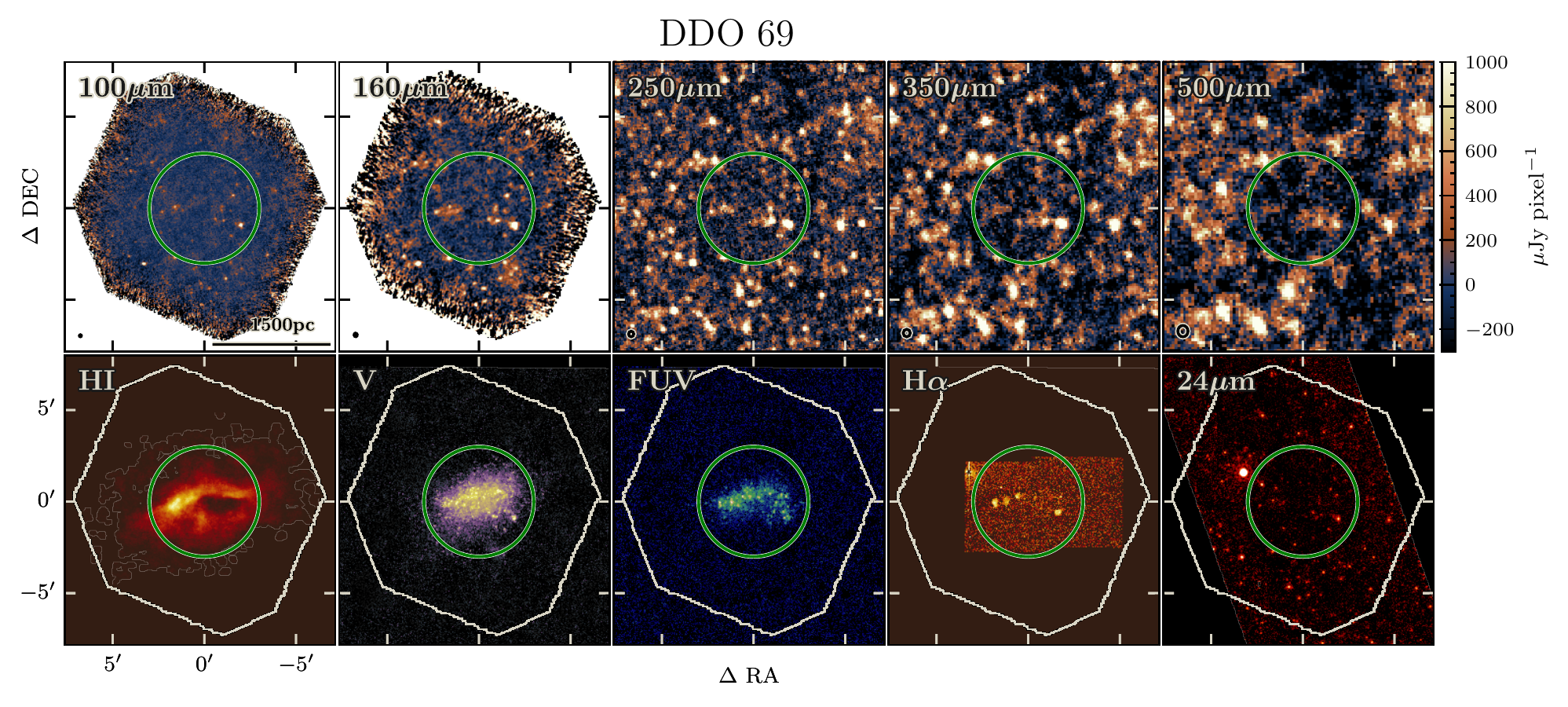}
	\\
    \includegraphics[width=1.0\textwidth]{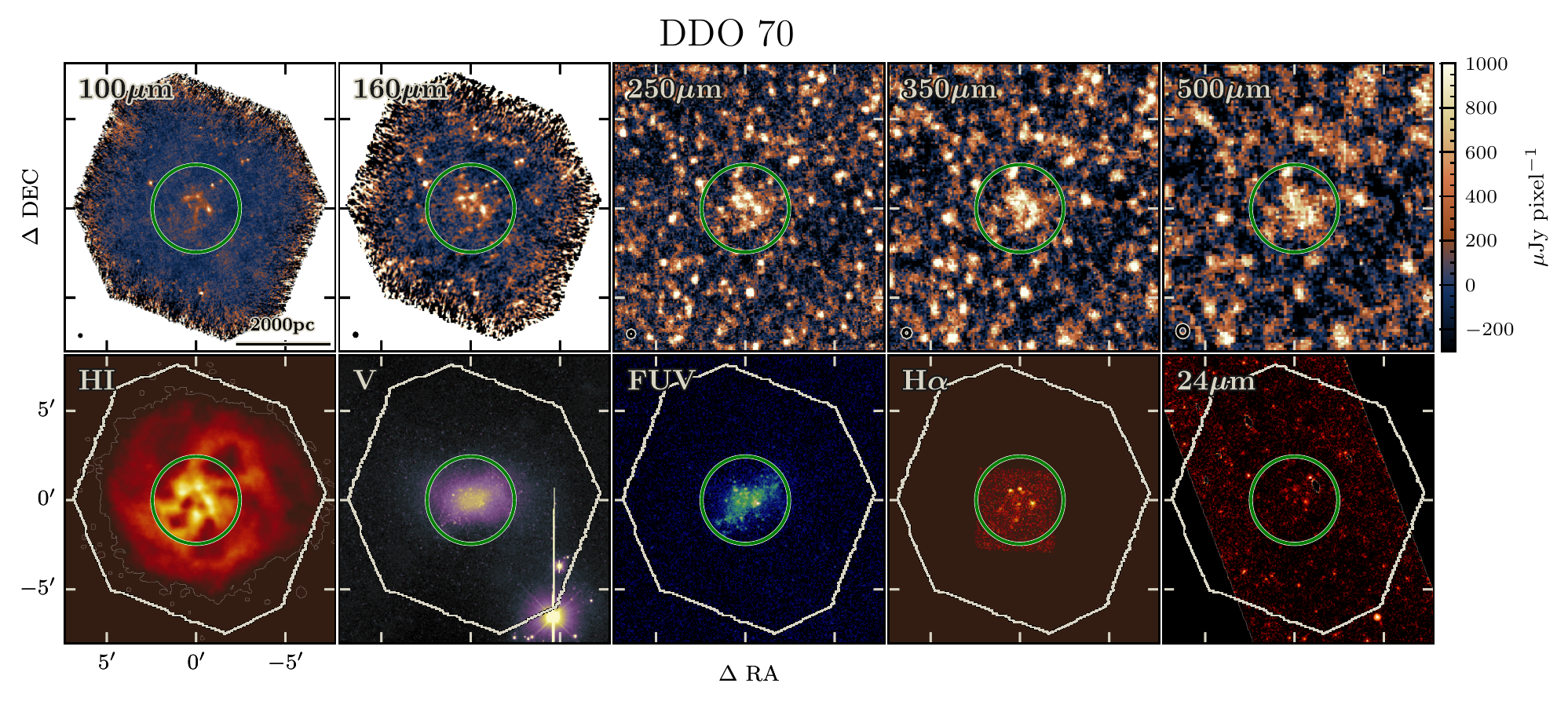}
\caption{Comparison of emission in the PACS and SPIRE maps with other observations at several wavelengths.
All PACS and SPIRE images are scaled from --0.3 to 1.0 mJy px$^{-1}$, and their respective beam sizes are represented by the black ellipses in the corners. The white outlines represent the extents of the PACS images, for scale.  The green circles denote the apertures used for photometry, as described in Table~\ref{table:Photometry} and \S~\ref{sec:apertureselection}.
}
\label{fig:mapcomp}
\end{figure*}

\begin{figure*}[ht!]
\centering
	\includegraphics[width=1.0\textwidth]{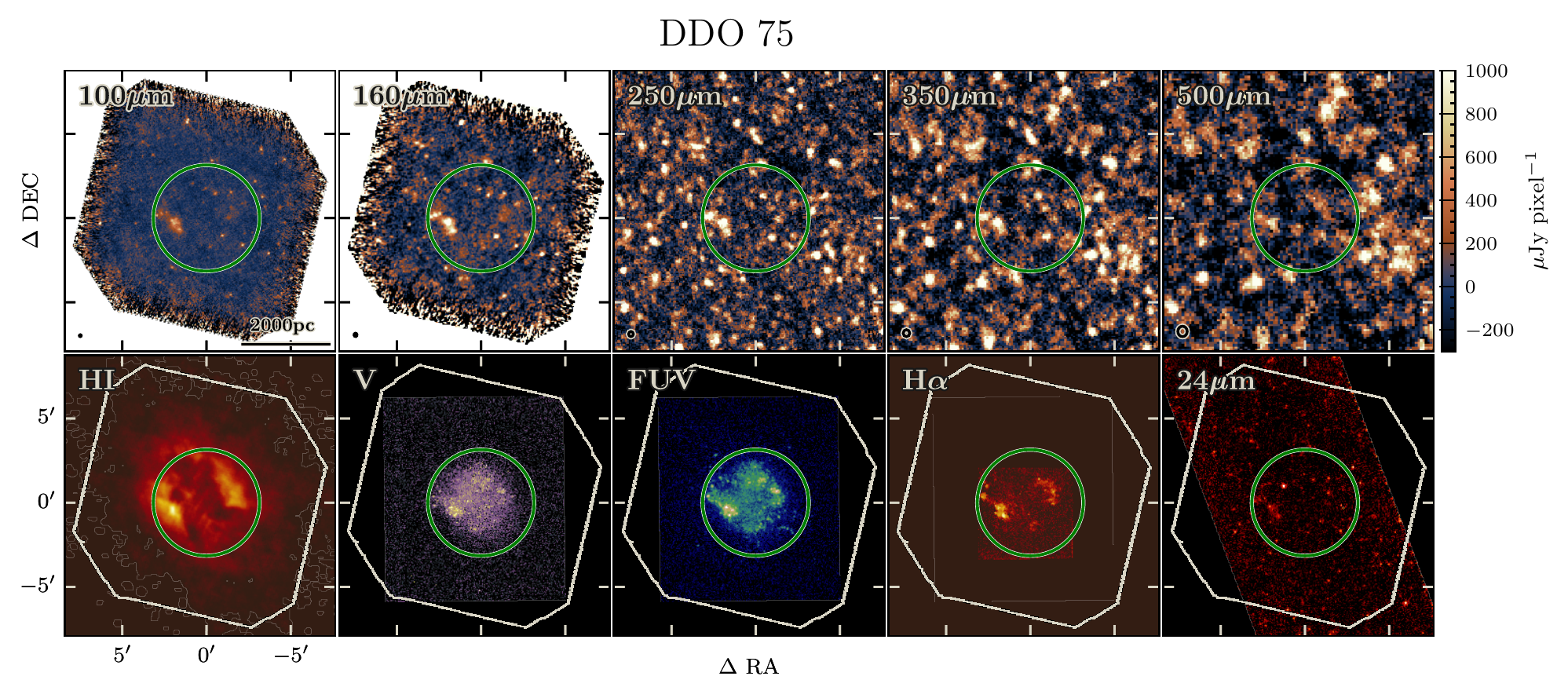}
	\\
	\includegraphics[width=1.0\textwidth]{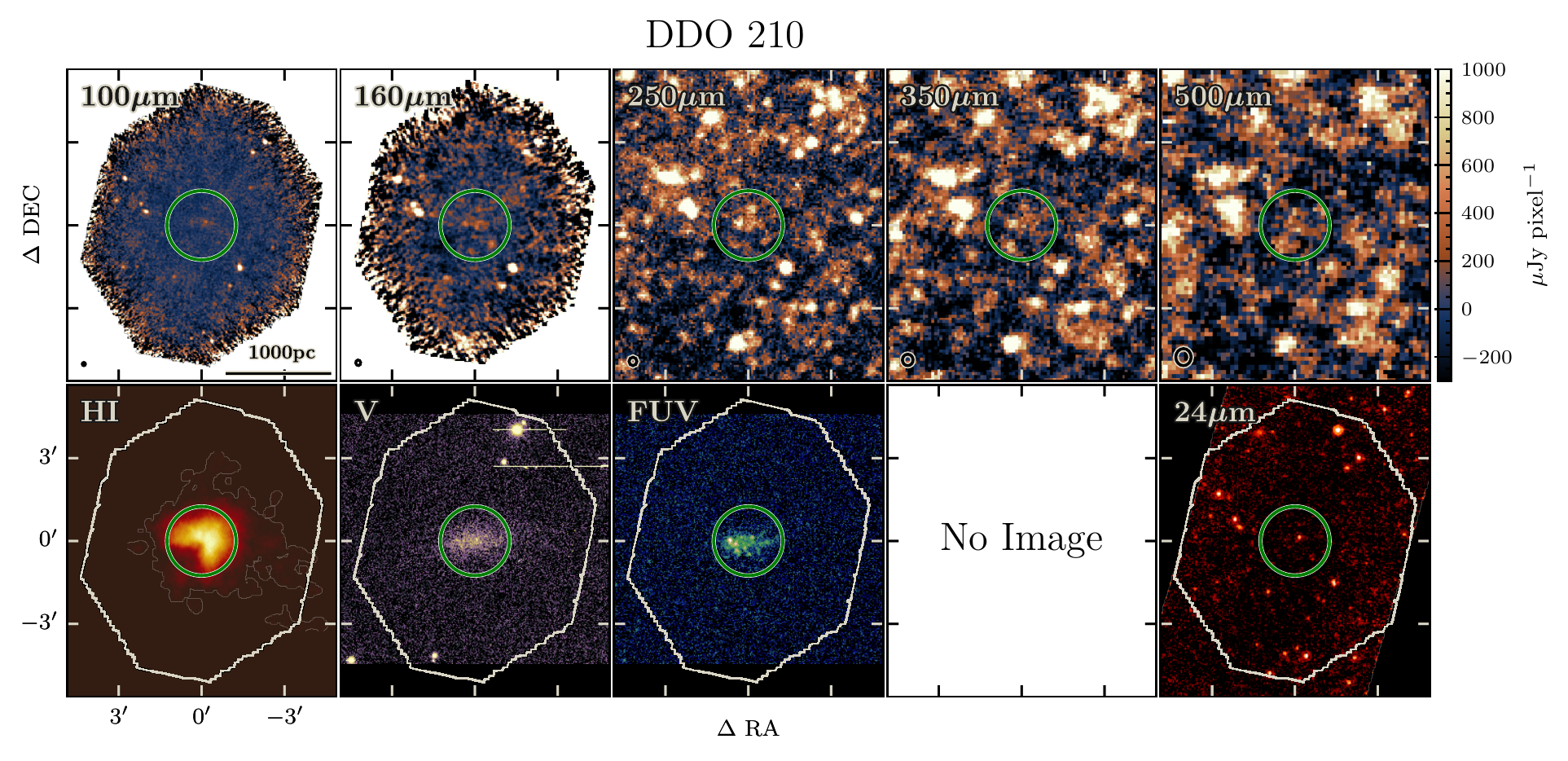}
\caption{Comparison of emission in the PACS and SPIRE maps with other observations at several wavelengths.
All PACS and SPIRE images are scaled from --0.3 to 1.0 mJy px$^{-1}$. The white outlines represent the extents of the PACS images, for scale.  The green circles denote the apertures used for photometry, as described in Table~\ref{table:Photometry} and \S~\ref{sec:apertureselection}.
}
\label{fig:mapcomp2}
\end{figure*}

We used the \herschel\ PACS \citep{Poglitsch2010} and SPIRE \citep{Griffin2010} instruments to observe our sources from 100--500$\mu$m. 
We refer the reader to the current PACS and SPIRE handbooks \footnote{\url{https://www.cosmos.esa.int/documents/12133/996891/PACS+Explanatory+Supplement}}\textsuperscript{,} \footnote{\url{https://www.cosmos.esa.int/documents/12133/1035800/The+Herschel+Explanatory+Supplement\%2C\%20Volume+IV+-+THE+SPECTRAL+AND+PHOTOMETRIC+IMAGING+RECEIVER+\%28SPIRE\%29}} for further information on the instrument characteristics as well as techniques and considerations regarding the processing of their data.
The PACS observations used scan-map mode at medium speed (20\arcsec\ s$^{-1}$).  The 100$\mu$m and 160$\mu$m bands were observed simultaneously over 2 repetitions to create maps of the sources.   

For SPIRE, the sources were also observed in ``large map'' mode at the nominal scan speed of 30\arcsec\ per second.  The 250$\mu$m, 350$\mu$m, and 500$\mu$m bands were all measured simultaneously over 4 repetitions.
Map sizes were set to 2$\times D_{25}$ from LEDA\footnote{\url{http://leda.univ-lyon1.fr/}} \citep{Makarov2014} on a side to ensure background could be determined beyond the measurable galaxian dust emission.  

A summary of the \herschel\ observations is provided in Table~\ref{table:Obs}.
The native full width at half maximum (FWHM) values for the observations are given in Table~\ref{table:instrumentparams}, and range from roughly 7\arcsec\ at 100\um\ to 38\arcsec\ at 500\um.
\spitzer\ mid infrared images of these galaxies at 24, 70, and 160\um\ were taken from the Local Volume Survey \citep{Dale2009}. 
Other ancillary data for much of the spectrum from UV to radio has been collected for each galaxy as part of the \lt\ survey, and in particular this work utilizes images of H$\alpha$ \citep{Hunter2004Ha}, $V$--band \citep{Hunter2006UBV}, FUV \citep{Hunter2010FUV}, and \hi\ \citep{Hunter2012LTdata} emission.
See also \citealp{Cigan2016} for more details.
We note here that the extents of the H$\alpha$ images are relatively small, as compared to both the images at other wavelengths and to the photometry apertures described in \S~\ref{sec:apertureselection}.

\vspace{1.0cm}

\subsection{Reduction}
\label{sec:reduction}

Basic data reduction was performed using the \herschel\ Interactive Processing Environment v12.1.0 \citep[HIPE;][]{Ott2010} with calibration trees 65 and 13.1 for PACS and SPIRE, respectively. In brief, for the PACS data, the raw instrumental signals were loaded at this stage, and preliminary processing was performed with the pipeline: the addition of pointing information, flagging of saturated pixels, conversion to flux density units (Jy pix$^{-1}$), flat-field calibration, electrical crosstalk corrections, and deglitching.  The resulting calibrated scan legs are called ``Level 1'' data products in the HIPE parlance.  The final steps required to make maps are to remove the bolometer drift noise and stitch together the many scan legs.

Final flux maps were produced using \textsc{Scanamorphos} \citep{Scanamorphos} v24.0.  \textsc{Scanamorphos} has been shown to be particularly effective at recovering faint extended flux in dwarf galaxies, as compared to the \textsc{MADmap} and \textsc{PhotProject} map-making methods \citep[see][]{RR13}.  \textsc{Scanamorphos} addresses low-frequency bolometer drift ``1/$f$'' noise by taking advantage of observation redundancy instead of assuming a noise model.   
The final maps were constructed with 1\farcs7 pixel sizes at 100$\mu$m and 2\farcs85 at 160$\mu$m, achieving finer sampling than Nyquist.  
The beam sizes for the different \herschel\ bands are listed in Table~\ref{table:instrumentparams}. 

For SPIRE, the HIPE Large Map pipeline was used to produce Level 1 data products, which involves adding pointing data, electrical crosstalk corrections for the bolometer arrays, deglitching, filter response calibration,  flux conversion (to Jy beam$^{-1}$), and time response correction.  
The final maps were produced with \textsc{Scanamorphos}, with uniform 4\farcs50, 6\farcs25, and 9\arcsec\ pixel sizes for all maps in the 250$\mu$m, 350$\mu$m, and 500$\mu$m bands, respectively.  See Figures~\ref{fig:mapcomp} and~\ref{fig:mapcomp2} for a comparison of the emission in several wavebands for each galaxy. 
Following the default procedure, relative gains (to account for beam area differences between bolometers) were applied in \textsc{Scanamorphos} instead of HIPE.

It should be noted that comparison of results from images produced with different map-making tools -- such as \textsc{Scanamorphos} and those within HIPE -- can have slight systematic differences.  
Specifically for this work, the PSFs of SPIRE maps produced by \textsc{Scanamorphos} are slightly modified due to additional smoothing\footnote{\url{https://nhscsci.ipac.caltech.edu//spire/docs/SPIRE_Mapmaking_Report_v6.pdf}}.  
\textsc{Scanamorphos} maps can also have smaller pixel sizes than those made from other tools, and do not contain blank pixels in their centers.

\vspace{2.5cm}

\section{PHOTOMETRY}
\label{sec:photometry}

\begin{figure*}[]
\centering
\includegraphics[trim=0mm 12mm 0mm 12mm, clip=true,width=1.0\textwidth]{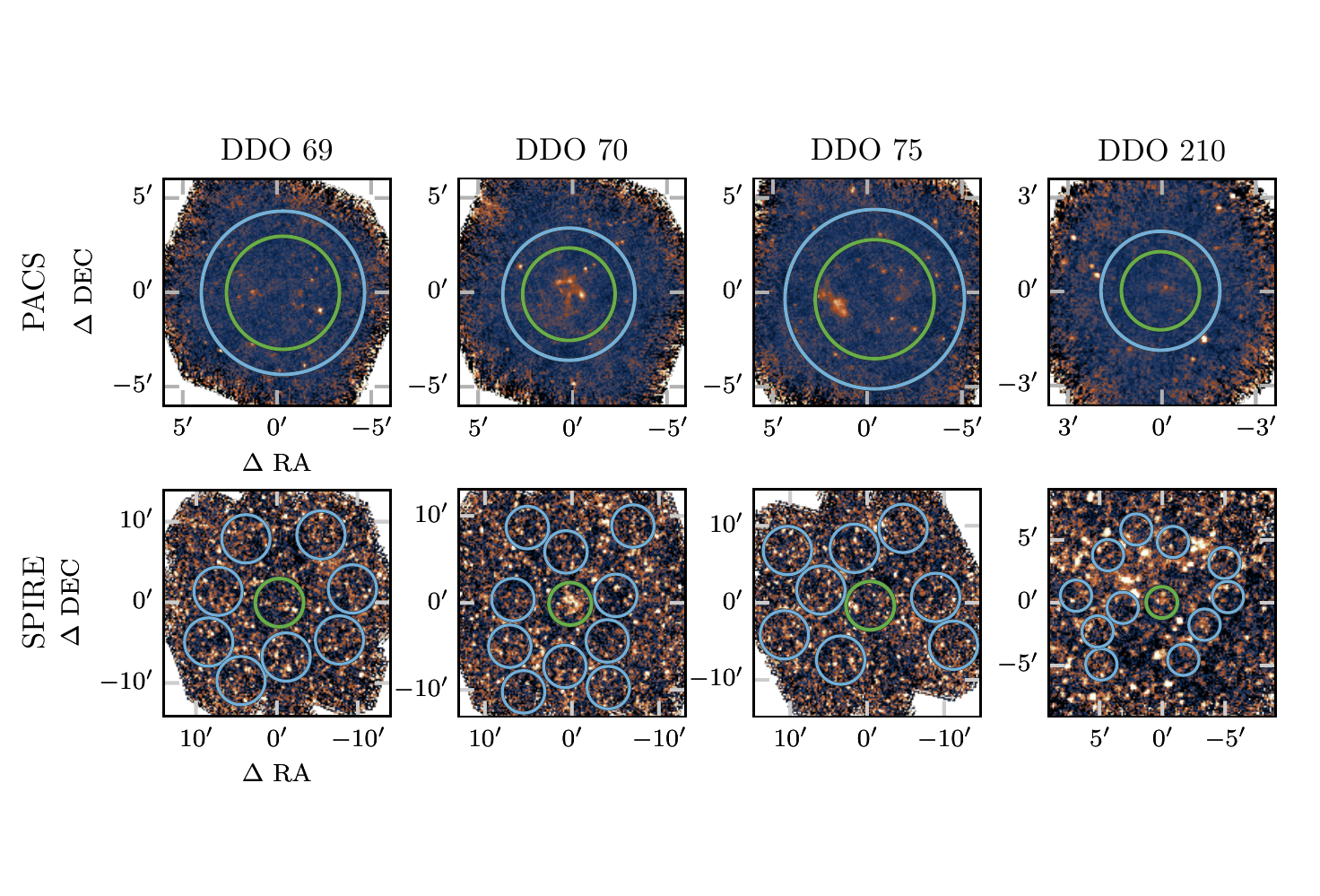}
\caption{Source and background photometry apertures used to derive the fluxes for each galaxy.  The source apertures, shown in green, are the same size between the PACS and SPIRE maps for a given galaxy.  The background apertures (annuli for PACS, circles for SPIRE) are shown in blue.  
All the PACS 100\um\ and SPIRE 250\um\ maps are plotted here with the same flux density scale, from $-0.3$ to $1.0 \ \textrm{mJy px}^{-1}$, also matching that in Figures \ref{fig:mapcomp} and \ref{fig:mapcomp2}.
}
\label{fig:photometryapertures}
\end{figure*}

\begin{deluxetable}{ llcc }[]
\tablecaption{ \herschel\ and \spitzer\ Instrument Parameters \label{table:instrumentparams}}
\tablehead{ \colhead{Instrument} & \colhead{Band} & \colhead{Beam Size} & \colhead{Calibration} \\
                                 &                &  (FWHM, arcsec) &  Uncertainty             }

\startdata
PACS  & 100$\mu$m  & 7.0 $\times$ 7.4    & 5\% \\
PACS  & 160$\mu$m  & 10.5 $\times$ 12.3  & 5\% \\
\\
SPIRE & 250$\mu$m  & 18.9 $\times$ 18.0 & 4+1.5\% \\
SPIRE & 350$\mu$m  & 25.8 $\times$ 24.6 & 4+1.5\% \\
SPIRE & 500$\mu$m  & 38.3 $\times$ 35.2 & 4+1.5\% \\
\\
MIPS &  24$\mu$m & 6.5  & 4\% \\
MIPS &  70$\mu$m & 18.7 & 10\% \\
MIPS & 160$\mu$m & 38.8 & 12\% \\
\enddata
\tablecomments{ The two linearly-additive calibration uncertainties listed for SPIRE include the 4\% absolute uncertainty in the model of Neptune and the 1.5\% random uncertainty in Neptune photometry measurements. }
\tablerefs{ Beam sizes come from the PACS and SPIRE Observer's Manuals.  Instrumental calibration uncertainties are as follows.  PACS: The PACS Observer's Manual; 
SPIRE: \cite{Bendo2013}; MIPS 24$\mu$m: \cite{Engelbracht2007}; MIPS 70$\mu$m: \cite{Gordon2007}; MIPS 160$\mu$m: \cite{Stansberry2007}.  }
\end{deluxetable}

\subsection{Aperture Selection}
\label{sec:apertureselection}

We select our source apertures by inspecting the emission in various bands from the optical to the infrared, to ensure that all of the galaxian emission is enclosed while still having space outside for background subtraction.  
The \hi\ images show that the distribution of the ISM is not always well-predicted by the optical emission, so the aperture choices should not be driven by optical emission alone. 
We use the same source aperture for every PACS and SPIRE band for a given galaxy, as shown in Figures~\ref{fig:mapcomp} and \ref{fig:mapcomp2}, so that flux levels are determined from equivalent areas.  
Circular apertures are used throughout, for careful treatment of aperture size effects and background subtraction, described more fully in the following sections. 
The aperture centers are the same as those used for the optical wavelength photometry of each galaxy, described by \cite{Hunter2006UBV}.  The radii $r$ are based on emission features from the ancillary data as follows:

\textit{DDO 69}: $r$=3\farcm0. 
This encompasses all of the bright emission that appears in the optical and FUV images.  This includes the slightly extended features in the PACS maps that correspond with H$\alpha$ emission, but excludes the bright H$\alpha$ spot to the northeast which appears to be a foreground star. The diffuse PACS 160\um\ emission near the southeast edge is excluded because it has no H$\alpha$, optical, or FUV counterpart.

\textit{DDO 70}: $r$=2\farcm45.  
This includes the bulk of the FUV and optical emission, and all of the H$\alpha$, while omitting the obvious compact FIR enhancements that have no optical or FUV counterparts. It is interesting to note that the apertures selected for the other three galaxies cover the majority of the \hi\ emission of their hosts, but this aperture for DDO 70 does not -- in fact, the \hi\ extends out to the noisy edge pixels of the PACS maps.  The bright spots in the PACS bands outside of our chosen aperture to the south, northeast, and northwest do not particularly correspond with any enhanced neutral hydrogen knots (at the \hi\ map resolution of $\sim$17\arcsec), whereas the emission inside the aperture does, so we treat the external spots as background sources.

\textit{DDO 75}: $r$=3\farcm15.  
This aperture encloses all of the H$\alpha$ emission, and all of the $V$--band and FUV except for a single bright spot 2\arcmin\ southeast of the edge in both of those images and some low-level emission just to the northeast in the FUV image.  This poses no problem, as the PACS maps have no corresponding emission in these locations.  The bright spot at the north edge of the chosen aperture has no optical or \hi\ counterpart, and is very faint at 24\um.

\textit{DDO 210}: $r$=1\farcm25. 
This aperture avoids almost all background sources and still covers essentially all of the emission at other wavelengths including \hi.  The bright features to the east and southwest of the defined aperture all appear at 24\um, but not in $V$--band or FUV, and they are clearly outside of the \hi\ extent of the galaxy.

\subsection{PACS Photometry}
\label{sec:pacsphotometry}

The careful calculations of the PACS flux densities and uncertainties described below build on the treatment described by \cite{RR13} and \cite{Ciesla2012}.

\subsubsection{Flux Extraction}
\label{sec:pacsfluxextraction}

Since the PACS photometer PSF has a persistent contribution out to around 1000\arcsec -- larger than the extents of our maps -- and since the background regions in the PACS maps are much closer to the source than in the SPIRE maps, the procedure for background level and source flux estimation are different from that of the SPIRE data.  In short, the background flux level cannot simply be computed as the simple median or mean of the pixel flux densities from the background regions because these contain a small contribution from the source (a few percent), which would cause the sky level to be overestimated.

Things are simplified if we use a circular source aperture and a concentric circular annulus for the background region.  The extent of the annulus in each map was chosen to be as large as possible without including large numbers of background sources, and without reaching the noisy pixels on the outer 1-2\arcmin\ of the maps (due to reduced coverage). The background regions used for each map are shown in Figure~\ref{fig:photometryapertures}.

We can measure uncorrected integrated flux densities within the source aperture and the annulus, labeled in lower case as $f_{\nu,\rm ap}$ and $f_{\nu,\rm ann}$, respectively.  Then, using the enclosed energy fractions ($\Phi$) of the PSF at each radius along with the number of integrated pixels in each aperture ($N_{\rm ap}$ and $N_{\rm ann}$), we can determine the true background level with the source contribution removed ($I_{\nu,\rm bg}$ Jy/pixel) and the final source flux $F_{\nu,\rm src}$ with a linear system of equations.

The raw flux density $f_{\nu,\rm ap}$ recovered from the source aperture is a fraction ($\Phi_{R0}$) of the total flux actually coming from the source, with the rest spread out over the 1000\arcsec\ PSF.  Furthermore, the background intensity $I_{\nu,\rm bg}$ for a given pixel, presumed to be uniform, is present on top of that.  The expression for the uncorrected source aperture flux can then be written as:

\begin{equation} \label{eq:f_ap}
f_{\nu,\rm ap} = \Phi_{R0} \cdot F_{\nu,\rm src} + N_{\rm ap} \cdot I_{\nu,\rm bg}.
\end{equation}

The raw flux summed in the annulus, $f_{\nu,\rm ann}$, would simply be $N_{\rm ann} \cdot I_{\nu,\rm bg}$ if there were no contribution except for sky emission.  However, the additional component from $F_{\nu,\rm src}$ is present, and the strength of this component depends on the fraction of PSF power enclosed between the inner radius $R1$ and outer radius $R2$.  That is, the additional source component in the background annulus is $\Phi_{R2} \cdot F_{\nu,\rm src} - \Phi_{R1} \cdot F_{\nu,\rm src} = (\Phi_{R2}-\Phi_{R1}) F_{\nu,\rm src}$.  Thus the raw integrated flux density in the background annular aperture is described by:

\begin{equation} \label{eq:f_ann}
f_{\nu,\rm ann} = N_{\rm ann} \cdot I_{\nu,\rm bg} + (\Phi_{R2}-\Phi_{R1}) F_{\nu,\rm src}.
\end{equation}

Solving these two equations with two unknowns, we obtain the solution for the corrected $F_{\nu,\rm src}$: 
\begin{equation} \label{eq:F_src}
F_{\nu,\rm src} = \frac{f_{\nu,\rm ann}-\frac{N_{\rm ann}}{N_{\rm ap}}f_{\nu,\rm ap}}{\Phi_{R2} - \Phi_{R1} - \Phi_{R0}\frac{N_{\rm ann}}{N_{\rm ap}}}.
\end{equation}

For a large source aperture radius of $R_0$=180\arcsec, the enclosed energy fraction $\Phi_{R0}$ is 0.976, which corresponds to a 2.4\% correction for the source aperture size at 100\um.  If the background annulus is $R_2$=220\arcsec\ and $R_1 = R_0$, then $\Phi_{R2}$ = 0.987 and the correction for source leakage into the background annulus is roughly 1.1\%.  The effect increases for our smallest source aperture of 75\arcsec, where the aperture size correction is 6.4\% at 100\um, and the leakage into a 115\arcsec\ background annulus is 4.3\%.
Formally, these aperture corrections are for single point source responses, and while they are good approximations for our maps which are not generally dominated by bright extended structure, the flux spread beyond the aperture could be slightly larger than what is accounted for here -- in particular for DDO 75, where there is extended emission near the aperture edge. 
This may be tempered somewhat by some background noise potentially being included in the contribution. 
We note that the background level could have been overestimated due to potential contamination from background sources which we do not estimate, so the source flux can be slightly higher, in principle.
These are relatively small corrections compared to uncertainties from, e.g., calibration, which will be discussed in the following section.
Color corrections, typically a few percent or less for these data, were applied during the fitting process (see \S~\ref{sec:SEDfit}) instead of at this stage, to better account for the shape of the FIR emission profile.

\vspace{0.8cm}

\subsubsection{Uncertainties}
\label{sec:pacsuncertainties}

The uncertainties $\Delta F_{\nu,\rm src}$ for the PACS map sums are calculated according to standard error propagation for Equation~\ref{eq:F_src}:  
\begin{equation} \label{eq:F_src_unc}
\Delta F_{\nu,\rm src} = \frac{\sqrt{\left(\Delta f_{\nu,\rm ann}\right)^2 + \left( \frac{N_{\rm ann}}{N_{\rm ap}} \ \Delta f_{\nu,\rm ap} \right)^2}}{\left| \Phi_{R2}-\Phi_{R1}-\Phi_{R0}\frac{N_{\rm ann}}{N_{\rm ap}}\right|} \ .
\end{equation}
We assume there is no uncertainty in the number of pixels or in the $\Phi$ values.
$\Delta f_{\nu,\rm ann}$ is a quadratic sum of two terms: the errors on the intensities of the individual pixels, and the uncertainty on the integrated background flux.  The individual pixel errors, from the ``error maps'' produced by \scanam, are summed as  $\sqrt{\sum \sigma^2_{i,\rm ann}}$.  The uncertainty on the summation of pixel intensities is taken to be the standard deviation of their values, $\sigma_{\rm sky}$, times $\sqrt{N_{\rm ann}}$.

$\Delta f_{\nu,\rm ap}$ is similarly composed of the individual pixel errors $\sigma_{i,\mathrm{ap}}$ and the uncertainty associated with summing intensities over the source aperture $\Delta_{\rm ap}$.  The latter would ideally be determined in the same manner as was done for the SPIRE data, from the variance of identical apertures placed at several independent positions in the image.  However, since the source aperture is generally too large to allow for other full apertures to fit inside the image, we can approximate the source flux uncertainty by using a smaller integration region to measure the variance, and scaling the result.  We use apertures reduced to $R_{\rm frac}=$ \sfrac{1}{2} the source aperture radius, allowing for 6--8 independent regions.   The number of pixels in an aperture goes as $R^2$, and uncertainty in a circular aperture typically goes as $\sqrt{N_{\rm pix}} = \sqrt{R^2} = R$ \citep[confirmed for PACS using radii above 100\arcsec\ by][]{Auld2013}, so we use $R_{\rm src}/R_{\rm frac}$ as the scaling fraction to obtain the final $\Delta_{\rm ap}$.  Tests on our maps, where full and reduced regions could be compared, verify that this scaling method yields appropriate results.

The uncertainties on the PACS annulus and source aperture measurements can be summarized as
\begin{equation} \label{eq:deltaf_ann}
\Delta f_{\nu,\rm ann} = \sqrt{\left(\sqrt{N_{\rm ann}}\sigma_{\rm sky}\right)^2 + \sum \sigma^2_{i,\rm ann} }
\end{equation}
and
\begin{equation} \label{eq:deltaf_ap}
\Delta f_{\nu,\rm ap} = \sqrt{\Delta_{\rm ap}^2 + \sum \sigma^2_{i,\rm ap} }
\end{equation}
for the determination of $\Delta F_{\nu,\rm src}$ 
with Equation~\ref{eq:F_src_unc}.  Again, the dominant contribution to the uncertainty is $\Delta_{\rm ap}$, responsible for $\sim$90\% of the non-systematic error and typically around 12\% of the measured flux.  The other components contribute one to a few percent each.

On top of these uncertainties, there is a systematic calibration uncertainty $\Delta_{\rm cal}$ which is 5\% of $F_{\nu,\rm src}$ in each of the PACS bands.  This is added in quadrature to $\Delta F_{\rm src}$, so that
\begin{equation} \label{eq:deltatot_PACS}
\Delta F_{\rm tot,PACS} = \sqrt{\left(\Delta{F_{\nu,\rm src}}\right)^2+\Delta^2_{\rm cal}}.
\end{equation}

\subsection{SPIRE Photometry}
\label{sec:spirephotometry}

\subsubsection{Flux Extraction}
\label{sec:spirefluxextraction}

The maps are first converted from units of Jy beam$^{-1}$ to Jy pix$^{-1}$, using the beam sizes listed in Table~\ref{table:instrumentparams} and the final pixel sizes of 4\farcs5 (250\um), 6\farcs25 (350\um), and 9\arcsec (500\um) as listed in $\S$~\ref{sec:reduction}.

Several (8--10) circular apertures surrounding the source region were defined in each map for the determination of the background flux level and aperture uncertainty. These background regions, shown in Figure~\ref{fig:photometryapertures}, were selected to fairly represent the sky level, avoiding areas dominated by high emission (background sources) and low emission levels alike.  
Our deep observations have notable contamination of background sources at the longer wavelengths, some of which undoubtedly fall within our source apertures.  To account for the background galaxy contribution in addition to the sky level, we take the median of the background aperture fluxes. 
The total source flux is obtained by summing the intensities within the source aperture, then subtracting the background value times the number of pixels in the source aperture.

The monochromatic flux densities produced by the pipeline are assumed to be for point sources.  The so-called ``$K_4$'' term from the SPIRE Observer's Manual is a frequency-weighted fraction of the Relative Spectral Response Function.  To correct for extended sources (diameter$>$FWHM$_{\rm beam}$), we multiply the integrated flux by a ratio of the $K_4$ factors for extended and point-like objects.  That is, $F_{\rm extended} = F_{\rm integrated} \times K_{4E}/K_{4P}$.  The ratios are 0.9986 at 250\um, 1.0015 at 350\um, and 0.9993 at 500\um.  This correction is required if the aperture is greater than 24\arcsec (250\um), 34\arcsec (350\um), or 45\arcsec (500\um), which applies to all of our source apertures.  Finally, an aperture correction is applied, dividing by the enclosed energy fraction of the PSF at the source aperture radius ($\Phi$).
As for PACS, color corrections are applied during the fitting procedure described in \S~\ref{sec:SEDfit}. 

To summarize:
\begin{equation} \label{eq:fluxtot_SPIRE}
F_{\nu,\rm tot} = \left(\sum_i^N I_{\nu,i} - N\cdot I_{\nu,\rm bg} \right)  \frac{K_{4E}}{K_{4P}} \, \frac{1}{\Phi}.
\end{equation}

\subsubsection{Uncertainties}
\label{sec:spireuncertainties}

The extracted SPIRE flux densities are subject to four types of non-systematic uncertainty.  Noise and flux integration effects related to the source aperture are estimated by $\Delta_{\rm ap}$.  This aperture uncertainty is calculated empirically as the standard deviation of the sums from each of the background apertures.  Since the background apertures are the same size as that of the souce, this provides a reliable estimate of the variation that can be expected from integrating fluxes over that area.  The determination of the background level contributes an uncertainty $\Delta_{\rm bg}$, calculated as the standard deviation of all background values, multiplied by $N_{\rm ap}/\sqrt{N_{\rm bg}}$.  The effect of individual uncertainties in each pixel's flux density from data reduction, summed in quadrature, is $\Delta_{\rm pix}$. Finally, the 4\% uncertainty in the beam area is $\Delta_{\rm beam}$.  We combine these four uncertainties in quadrature, so that the total uncertainty in the flux density determination is
\begin{equation} \label{eq:fluxunc_SPIRE}
\Delta_{\rm flux}^2 = \Delta_{\rm ap}^2 + \Delta_{\rm bg}^2 + \Delta_{\rm pix}^2 + \Delta_{\rm beam}^2.
\end{equation}
This could in practice be an overestimate of the true uncertainty, as the components may be slightly correlated. 

There is a systematic calibration uncertainty of 5.5\% applied to all SPIRE bands.  This is comprised of a 4\% absolute uncertainty in the modeled flux density of Neptune and a 1.5\% random component resulting from the repeated Neptune measurements. As advised in the SPIRE Observer's Manual, these calibration errors are added linearly instead of in quadrature.  Upper limits to the flux density ($3\sigma$) are reported for maps where the integrated flux is less than three times the non-systematic (``internal'') uncertainty.

The dominant source of uncertainty is $\Delta_{\rm ap}$.  For example, in the DDO 70 250\um\ measurement, $\Delta_{\rm ap}$ is 88.3 mJy, comprising about 90\% of the non-systematic error, or 15\% of the measured flux. 
$\Delta_{\rm bg}$ is the smallest component at 5 mJy, less than 1\% of the measured flux density.  $\Delta_{\rm pix}$, $\Delta_{\rm beam}$, and the calibration uncertainty are all similar in scale, around 20 mJy, contributing a few percent of the flux level each.  The final integrated flux densities and uncertainties for each map are tabulated in Table~\ref{table:Photometry}.

\subsection{Photometry of Smaller Regions}
\label{sec:miniregs}

\begin{figure*}[]
\centering
\includegraphics[width=1.0\textwidth]{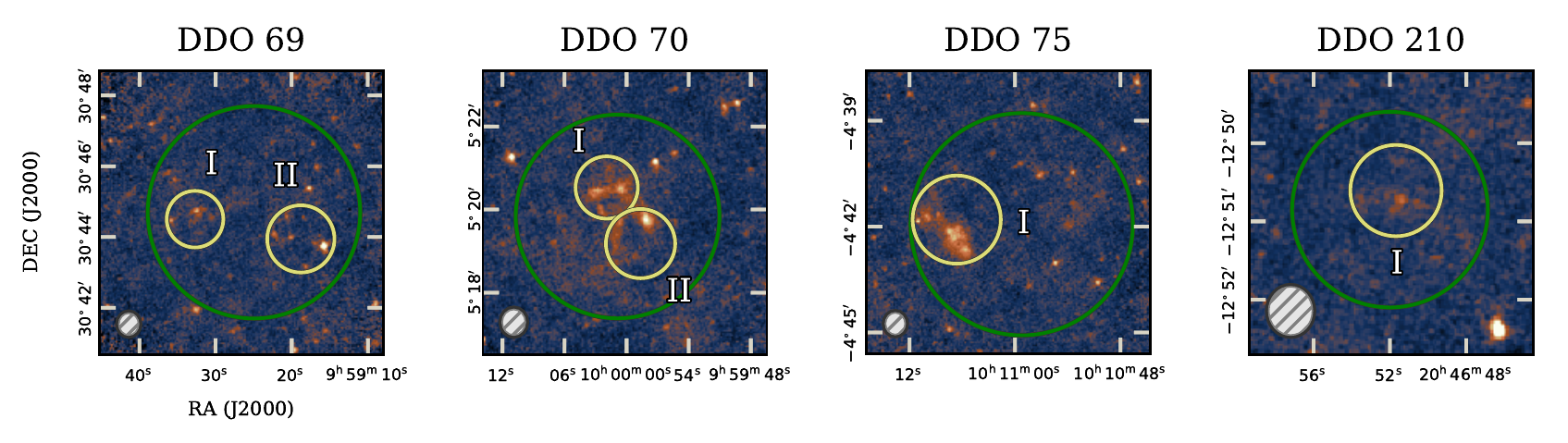}
\caption{The apertures used to study smaller regions of source emission in each galaxy are shown in yellow. The original whole-galaxy aperture is shown in green over the PACS 100\um\ map, for reference.  The largest beam size of the observations -- that of SPIRE 500\um\ -- is shown in the lower left corner of each map.
}
\label{fig:minireglocations}
\end{figure*}

The source of emission in dwarfs can be relatively compact compared to the full aperture that encompasses all the galaxy FIR emission, leaving large areas of negligible signal that can reduce the overall signal-to-noise ratio of the full-aperture integrated fluxes.  In order to draw out the maximal signal and study resolved regions within a particular galaxy, we perform additional aperture photometry over smaller regions within the full-galaxy apertures.  Flux densities were determined using the same procedures as for the full-sized apertures, except for the aperture uncertainty in the PACS images: with the smaller source apertures, equally-sized offset apertures can be used for the empirical estimate, so no scaling factor is necessary.  The regions used are displayed in Figure~\ref{fig:minireglocations}, and the results of the photometry are listed along with the full aperture results in Table~\ref{table:Photometry}.

We note that the PACS aperture uncertainties are expected to be more significant for these smaller apertures than for the full galaxy integrations.  
Analyzing the large 4\degr$\times$4\degr\ tiles of the \herschel\ Virgo Cluster Survey, \cite{Auld2013} found that the aperture uncertainty increases as the usual $N_{\rm pix}^{1/2}$ at large radii ($\gtrsim 100$\arcsec), but that it follows a steeper $N_{\rm pix}^{3/4}$ trend for smaller radii, with the turnover occurring at several times the beam FWHM.  
This is likely due to the large-scale structure of the PSF causing even relatively distant pixels to be partially correlated.

\begin{deluxetable*}{ lcrrccccccc }[t!]
\tabletypesize{\footnotesize}
\tablecaption{ Photometry \label{table:Photometry}} 
\tablehead{ \colhead{Galaxy} & \colhead{Region} & \multicolumn{2}{c}{Aperture Center} & \multicolumn{2}{c}{Aperture Radius} & \colhead{F$_{100}$} & \colhead{F$_{160}$} & \colhead{F$_{250}$} & \colhead{F$_{350}$} & \colhead{F$_{500}$}  \\
 & & \multicolumn{1}{c}{RA} & \multicolumn{1}{c}{DEC} & (\arcsec) & (pc) & (mJy) & (mJy) & (mJy) & (mJy) & (mJy) }

\startdata
DDO 69  & Whole & 9 59 25.0  & 30 44 42.0  & 180 & 698& $309 \pm 90$ & $336 \pm 87$ & $136 \pm 29$ & $\leq  101 $ & $\leq  116 $   \\ 
DDO 70  & Whole & 10 00  0.9  & 5 19 50.0  & 147 & 926& $1037 \pm 143$ & $862 \pm 125$ & $590 \pm 99$ & $358 \pm 72$ & $211 \pm 61$   \\ 
DDO 75  & Whole & 10 10 59.2  & -4 41 56.0  & 189 & 1191& $914 \pm 141$ & $890 \pm 157$ & $365 \pm 87$ & $155 \pm 88$ & $\leq  199 $   \\ 
DDO 210  & Whole & 20 46 52.0  & -12 50 51.0  & 75 & 327& $71 \pm 21$ & $103 \pm 28$ & $87 \pm 20$ & $\leq  68$ & $\leq  42$   \\ \\
\multicolumn{11}{c}{Subregions} \\
\hline 
DDO 69  & I  & 9 59 32.7 & 30 44 30.3  & 48  & 186 & $126 \pm 18$ & $168 \pm 24$ & $47 \pm 18$ & $\leq 41$ & $\leq  32 $   \\ 
 & II  & 9 59 18.8 & 30 43 56.9  & 56  & 221 & $154 \pm 30$ & $259 \pm 40$ & $191 \pm 28$ & $94 \pm 20$ & $33 \pm 15$   \\ 
DDO 70  & I  & 10 00  1.9 & 5 20 31.8  & 45  & 283 & $341 \pm 33$ & $301 \pm 37$ & $191 \pm 23$ & $111 \pm 16$ & $45 \pm 9$   \\ 
 & II  & 9 59 58.7 & 5 19 10.8  & 50  & 315 & $306 \pm 35$ & $243 \pm 35$ & $176 \pm 27$ & $109 \pm 18$ & $72 \pm 12$   \\ 
DDO 75  & I  & 10 11  6.6 & -4 41 48.1  & 75  & 472 & $577 \pm 36$ & $396 \pm 100$ & $241 \pm 46$ & $86 \pm 31$ & $24 \pm 23$   \\ 
DDO 210  & I  & 20 46 51.7 & -12 50 36.4  & 35  & 152 & $52 \pm 11$ & $59 \pm 16$ & $74 \pm 17$ & $28 \pm 11$ & $9 \pm 7$   \\ 
\enddata

\tablecomments{Integrated aperture flux densities for each observed band.  
Upper limits are given as 3 times the non-systematic uncertainty.  All other uncertainties include the contribution from systematic errors.  We note that the combined subregion flux densities in DDO 69 at 160 and 250\um\ exceed the full-galaxy values in those bands; this is likely due to the large number of noisy low and negative flux pixels in the full source aperture. 
These values do not contain color corrections, which were determined during the fitting process and are listed in Table~\ref{table:SEDs}.  
}

\end{deluxetable*}


\section{Comparison of Observed Bands}
\label{sec:Bandcomp}

\subsection{Comparison with \spitzer}
\label{sec:spitzercomp}

\begin{figure}[h]
\centering
\includegraphics[width=1.0\linewidth]{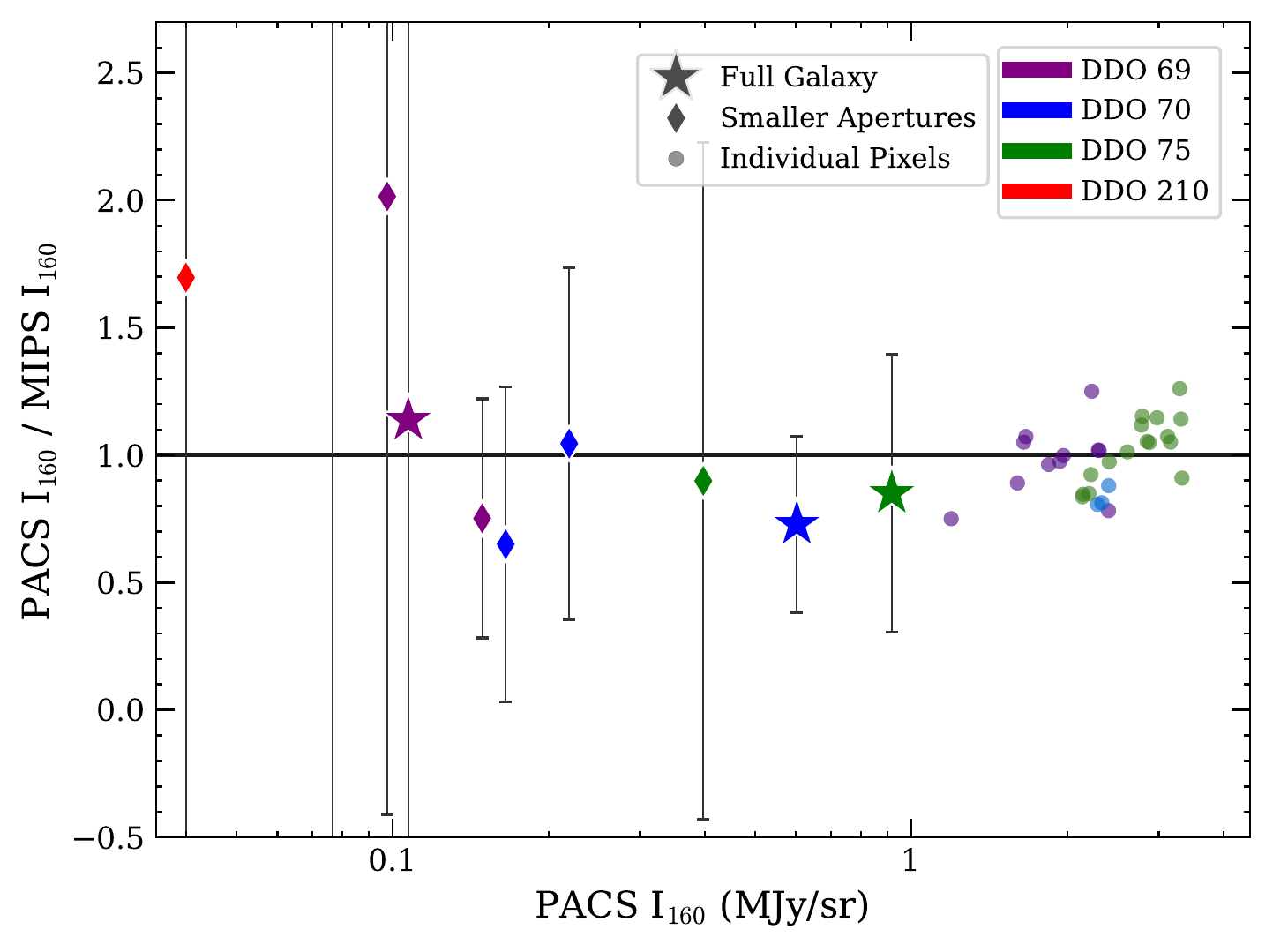}
\caption{Comparison of PACS 160\um\ and MIPS 160\um\ measurements.  
The horizontal line shows where the fluxes are equal between the instruments.
The full-galaxy and smaller aperture values (stars and diamonds, respectively) are consistent within the errors. 
The individual pixels above three times the rms level in both PACS and MIPS maps (circles) also show some scatter, but fall near the line of equivalence. 
The brightest pixels have a slight prevalence for PACS/MIPS ratios $>$ 1, with ratios in the fainter pixels trending to slightly below parity. 
}
\label{fig:PACSvsMIPS}
\end{figure}

It is important to compare measurements from different instruments in the same wavebands, as this gives an impression of the reliability of the data.  PACS and MIPS both have bands at 70\um\ and 160\um, however the \lt\ sample only has common data for both at 160\um.  The MIPS data for all four galaxies discussed in this photometry study were presented by \cite{Dale2009} as part of the \spitzer\ Local Volume Legacy
Survey.  MIPS fluxes were integrated using the same apertures as for the PACS measurements to make this comparison, after the PACS maps were convolved to the MIPS beam size and reprojected to an identical pixel grid.  Figure~\ref{fig:PACSvsMIPS} shows these ratios of PACS$_{160}$/MIPS$_{160}$ surface brightnesses for the \lt\ sample, where ratios are given for the full galaxy integrations, the smaller apertures, and on a pixel-by-pixel basis. 

These objects are all faint, with low absolute fluxes compared to brighter galaxies.  This means that the same absolute offset in fluxes for inherently faint targets may yield deceptively high or low ratios compared to brighter, high S/N sources.  DDO 75, the brightest in this sample, has a PACS/MIPS ratio of 0.85$\pm$0.54 -- consistent within the error.  
This mirrors the trend noted in the spectroscopic study of \lt\ dwarf galaxies \citep{Cigan2016} where the total infrared luminosity determined from the \cite{Galametz2013} prescription for PACS 100\um\ was typically about half that determined from MIPS 24, 70, and 160\um.
DDO 69 is much fainter overall, and the PACS flux for the full galaxy is 1.4 times higher than that determined from the notably noisy MIPS image. 
DDO 210 is fainter yet, with a small negative integrated flux from the MIPS 160\um\ map combined with a large uncertainty: $-0.01 \pm 0.14$ Jy. Indeed, no detection is obvious upon manual inspection of the MIPS image. Thus dividing the PACS value by these results gives a very large negative ratio and an even larger error.  These large differences occur because the whole-galaxy apertures for these two systems include many pixels that are essentially just noise. 
The smaller photometry regions can vary between instruments by factors of up to 2, though the average of all their PACS/MIPS intensity ratios is 1.01, in line with unity.

Individual bright pixels (those greater than three times the rms level in each of their respective maps) show some scatter between the two instruments, but are consistent with parity overall.  
The brightest PACS pixels appear to be slightly brighter in PACS than in MIPS by a factor of $\sim1.2$ on average, slowly decreasing to become fainter in PACS by the same factor as they near the rms level, though the scatter is several times that difference. 
The pixel values below the rms level are consistent with random noise, and the noise pixels in PACS generally correspond to noise pixels in MIPS. 
The PACS Handbook provides correction factors for comparing PACS densities with monochromatic flux densities from MIPS, based on each instrument's bandpasses, for a variety of source modified blackbody profiles (discussed more in the following sections; see Eq.~\ref{eq:modifiedBB}). 
For typical values expected in galaxies of temperatures in the range of 15--30 K, and $\beta$ between 1--2, these correction factors range from 0.95--1.05. 
This is smaller than the scatter seen in the bright pixels and integrated fluxes in Fig.~\ref{fig:PACSvsMIPS}, so the variation cannot be attributed entirely to differences in bandpass response. 

The bandpass overlap between MIPS 70\um\ and PACS 100\um\ is insufficient for any meaningful comparison, so no attempt was made for this pair. 
Although MIPS 70\um\ maps exist for the \lt\ sample, they contain significant noise in stripes that are difficult to cleanly remove from the images, so we do not consider them for the remainder of this work.

\subsection{\herschel\ FIR Colors}
\label{sec:colors}

{
\begin{figure*}[]
\centering
	\includegraphics[width=.8\linewidth]{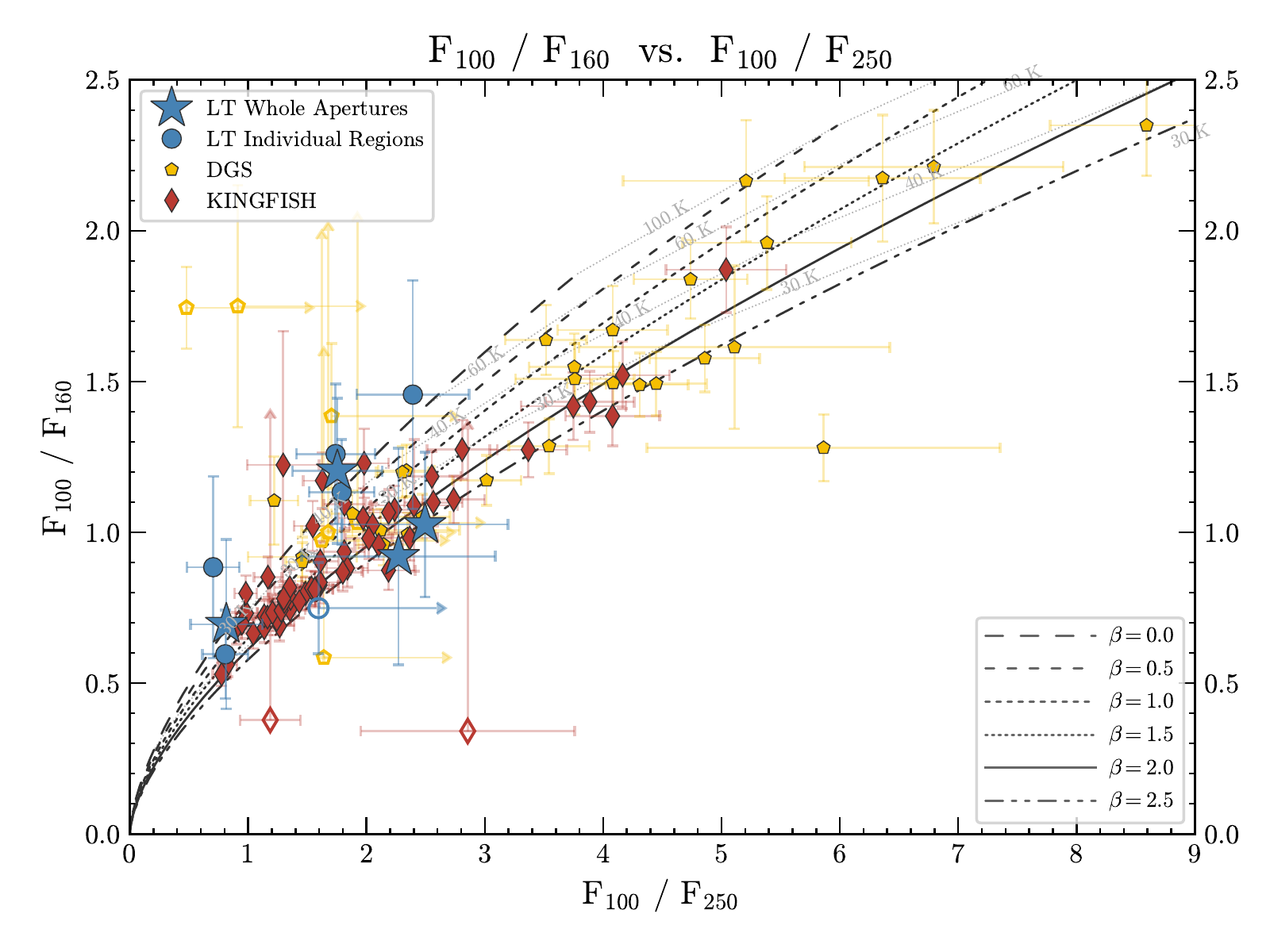}
	\\
	\includegraphics[width=.8\linewidth]{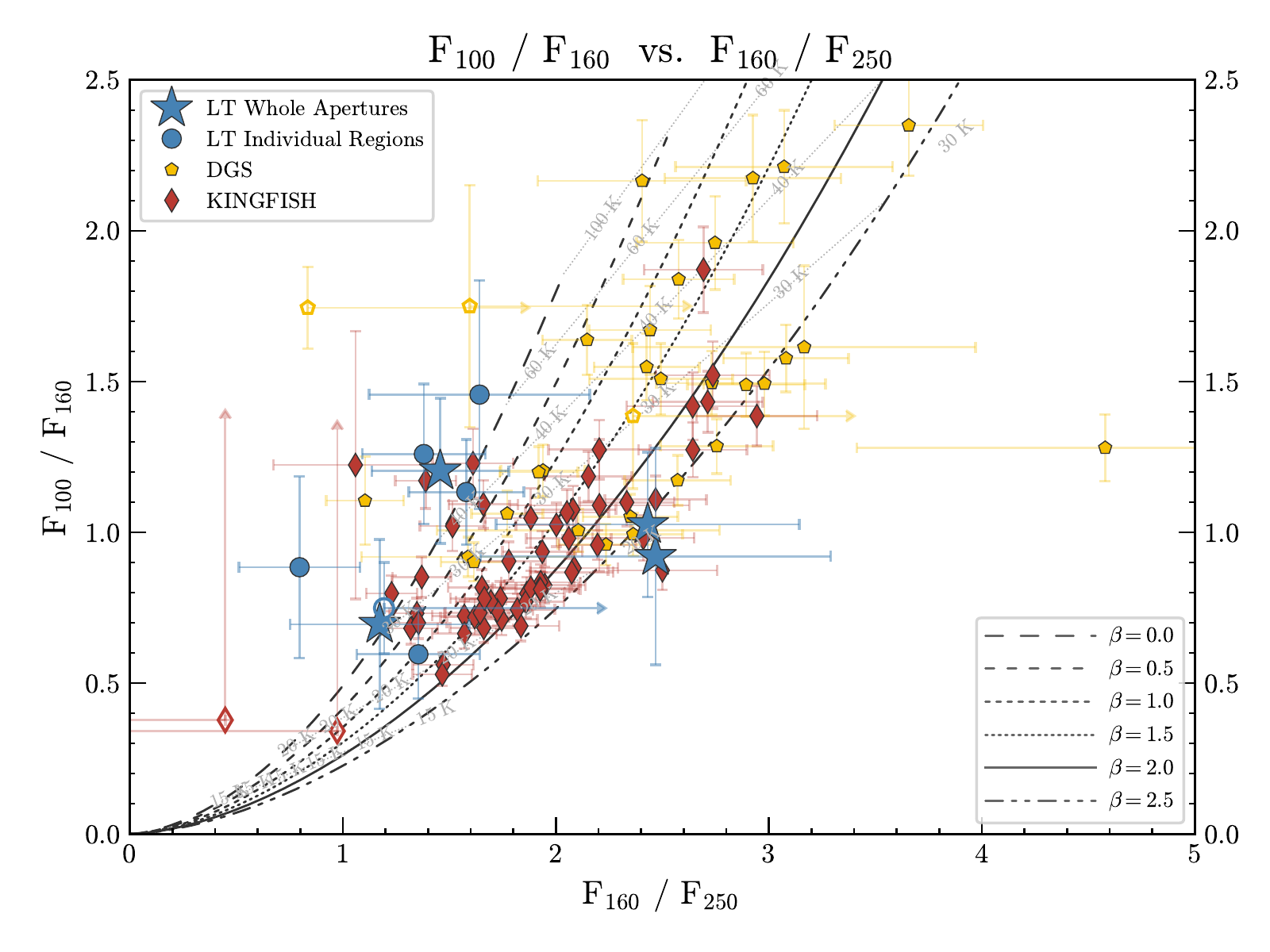}
\caption{FIR color-color diagrams for the DGS and KINGFISH galaxies, with theoretical modified blackbody curves at fixed T and $\beta$ values, as presented by \cite{RR13} but using the updated fluxes from \cite{RemyRuyer2015}.  
\lt\ detections are denoted as solid stars (full-galaxy apertures) and circles (small targeted regions), while the open symbols are limits. 
}
\label{fig:colorcolor1}
\end{figure*}

\begin{figure*}[]
\centering
	\includegraphics[width=.8\textwidth]{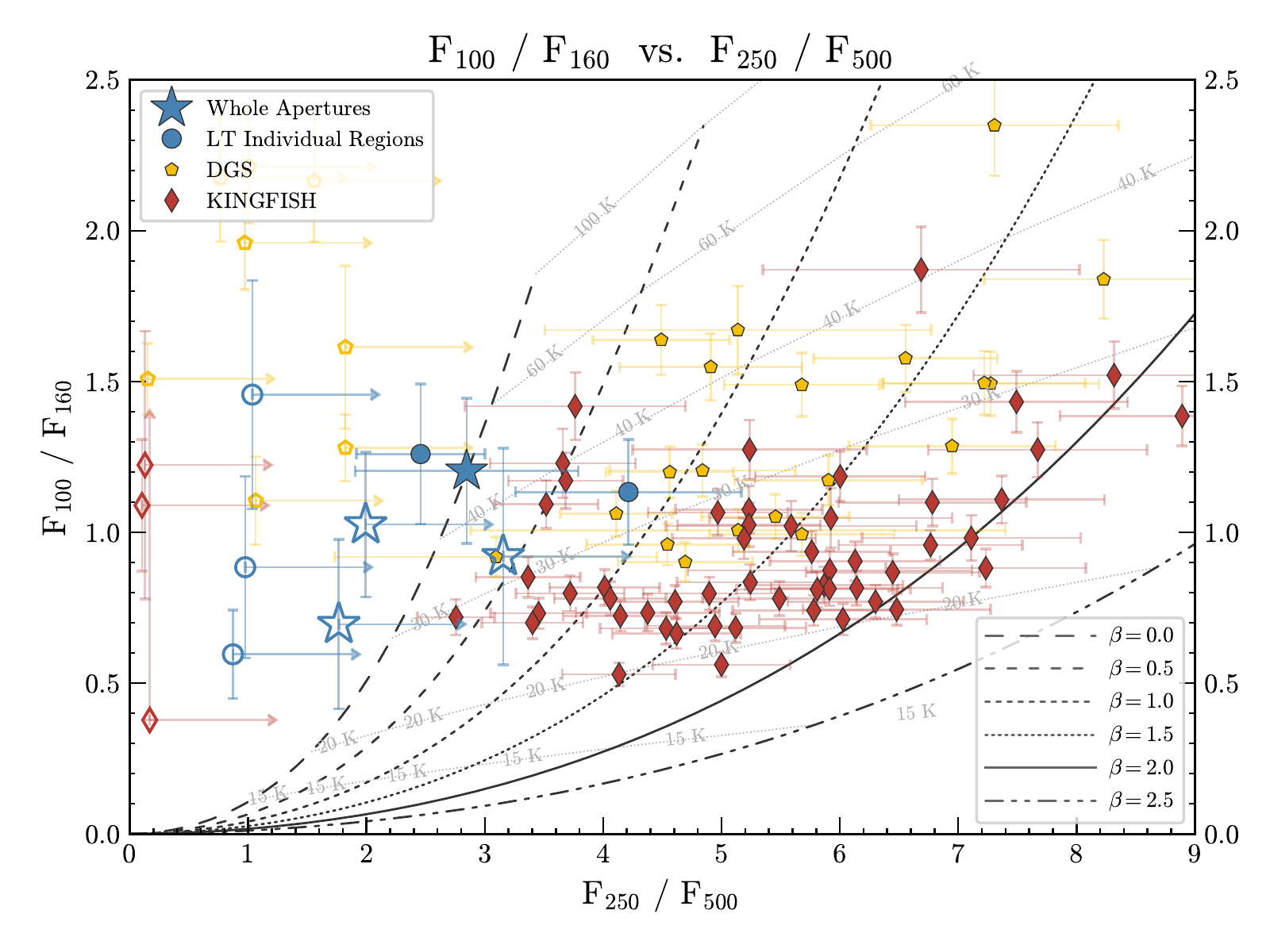}
	\\
	\includegraphics[width=.8\textwidth]{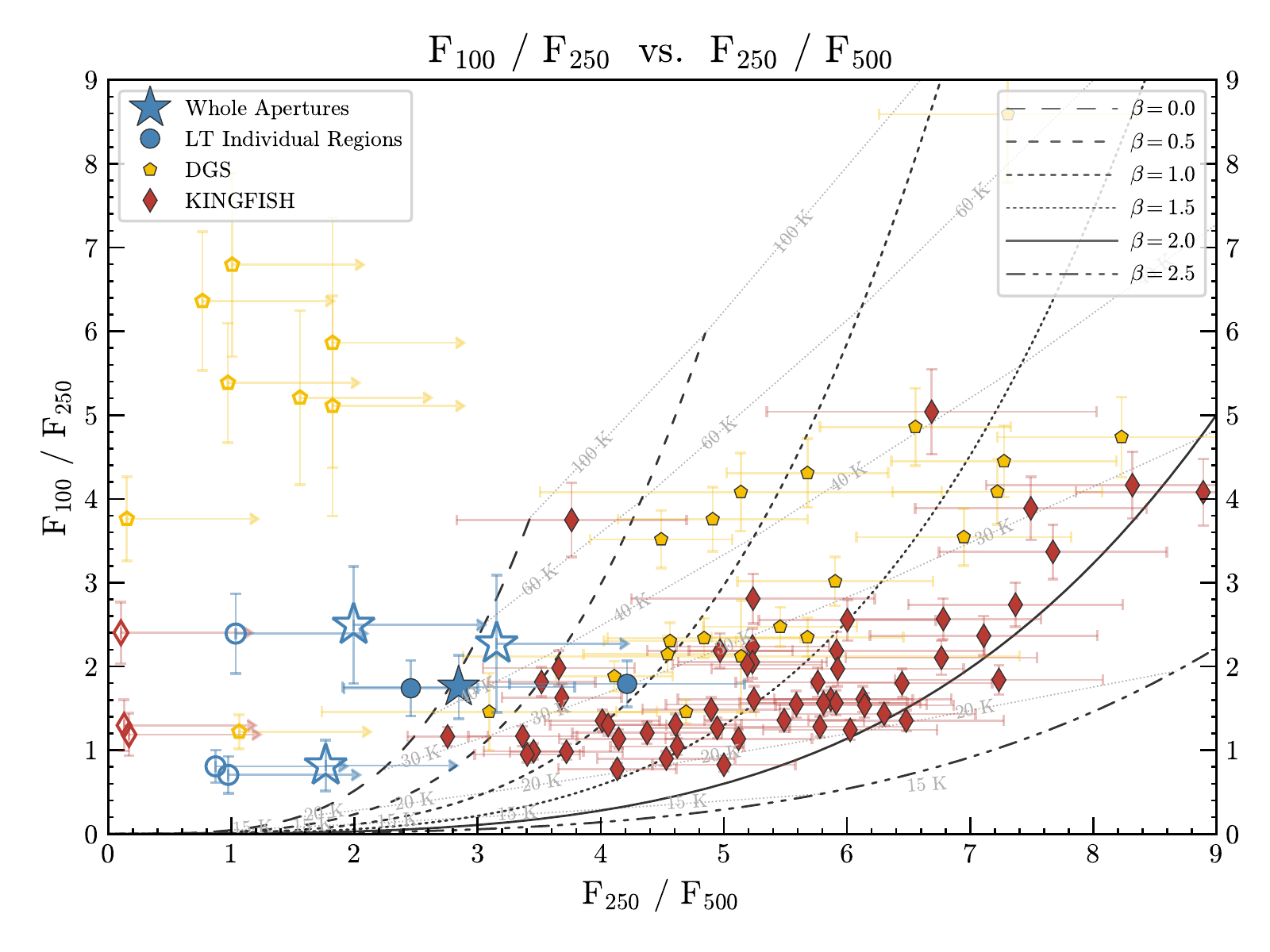}
\caption{ Color-Color diagrams at FIR and sub-mm wavelengths, following the treatment of \cite{RR13}. 
The symbols are the same as in Figure~\ref{fig:colorcolor1}.
}
\label{fig:colorcolor2}
\end{figure*}

Color-color diagrams give a qualitative view of the distribution of IR light in the observed galaxies by relating the relative fluxes in two different wavebands.  This can also be thought of in terms of the SED.  For bands near the peak of a blackbody curve, for example, different colors (relative heights on the SED) could potentially be a reflection of information such as the temperature of the dust.   The modified blackbody model which is commonly used to describe the FIR continuum emission in galaxies (discussed in more detail in \S~\ref{sec:SEDmodel}), depends non-linearly on temperature $T$ and dust emissivity $\beta$. It scales as $F(\lambda)\propto \lambda^{-\beta} B(\lambda,T)$ where $B$ is the Planck function, assuming a simple single temperature model of dust with a specific set of optical properties.

All of the \lt\ galaxies were detected in the PACS bands, but only DDO 70 was detected at 3$\sigma$ significance in all three SPIRE bands.  Reducing the aperture size to the targeted subregions improves the signal to noise ratio due to the exclusion of many pixels without substantial emission.  In Figures \ref{fig:colorcolor1} and \ref{fig:colorcolor2}, comparisons of various FIR colors are shown for the \lt\ galaxies, along with the DGS and KINGFISH \citep{KINGFISH} colors from the fluxes reported by \cite{RemyRuyer2015} \citep[slightly updated from those in][]{RR13}.  Theoretical curves for blackbodies of varying temperatures and emissivities are also plotted.  

The KINGFISH galaxies are generally more tightly clustered than the DGS galaxies, which spread across a large area of parameter space.  \cite{RR13} attribute this to differences in the dust properties between the samples, as the KINGFISH galaxies tend to be more metal-rich.  
The \lt\ sample values fall among the spread seen in the other samples. 

Since FIR SEDs of most of these galaxies peak between 100--160\um, the 100--160 and 160--250 colors can allow us to study the cold ISM properties between samples. Flux densities that follow $F_{100}>F_{160}>F_{250}$ would indicate SEDs that peak shortward of 100\um.  
The KINGFISH galaxies are clustered in the region depicting cooler dust, while the DGS galaxies show colors more typical of warmer dust as shown by \cite{RR13}.
The \lt\ galaxies fall in the mid to low range, indicating cooler dust than the bulk of the DGS galaxies, with some of the \lt\ regions having very cold dust.

There are several caveats and important factors to note when interpreting these diagrams.  Data do sometimes fall outside of the predicted ranges -- the outliers in the DGS sample were all noted to be faint, low-metallicity galaxies, and may be due to the fact that a single temperature and $\beta$ is not always a realistic approximation.  Each region should likely have a distribution of T and $\beta$ (emissivity) values.   
Generally, larger $\beta$ can be countered by smaller T to produce the same color, and the degeneracy can be quite pronounced, as seen in the top panel of Figure~\ref{fig:colorcolor1}.   
The SPIRE bands are more likely to sample the Rayleigh-Jeans limit of a blackbody, as opposed to the peak of the SED for warm dust or stochastically heated small grain emission which occurs at shorter wavelengths ($<24$\,$\mu$m).  Flatter slopes, or broad SED peaks, can have several causes.   One explanation is low $\beta$ values in the dust.  However, this could be mimicked by a wide distribution of dust temperatures in the source.  

Many models assume that $\beta$ should be fixed or constrained to be near a value of 2, as thought to be typical of grains in the Milky Way \citep{planckXXIX}.  If $\beta$ is assumed to be 2, low 250/500\um\ ratios can be interpreted as additional emission in the long SPIRE bands (the so-called ``sub-mm excess'' e.g., \citealp{Galliano2003, Grossi2010,Galametz2011,Draine2012}), beyond what would be expected from a single blackbody.  This expectation of $\beta=2$ grains in the low-metallicity environments of dwarfs could simply be flawed, however.  Small $\beta$ values and slightly higher ratios of 500\um\ to shorter wavelength fluxes can also indicate enhanced emission longward of 500\um, as was found in several DGS dwarfs. 
Although the 500\um\ detections in DDO 70 are consistent with small $\beta$ values, the longest-wavelength flux densities are well-fitted by a single blackbody curve (discussed in \S~\ref{sec:SEDs}), with no hint of elevated sub-mm emission.


\section{Determination of FIR Continuum and Dust Properties}
\label{sec:SEDs}

\subsection{SED Model}
\label{sec:SEDmodel}

To derive the properties of dust in our sample of four dwarf galaxies, we approximate the FIR emission as a modified blackbody function based on three free parameters: temperature ($T$), dust mass ($M_{\rm dust}$), and the dust emissivity index ($\beta$). The form of the function is as follows:
\begin{equation} \label{eq:modifiedBB}
F_\nu(\lambda) = \frac{M_{\rm dust} \, \kappa(\lambda_0)}{D^2} \left(\frac{\lambda}{\lambda_0}\right)^{-\beta} B_\nu(\lambda,T).
\end{equation}
Here, $B_\nu(\lambda,T)$ is the Planck function, $D$ is the distance to the source (given in Table~\ref{table:sampledata}), $\lambda_0$ is the reference wavelength of 160\um, and $\kappa(\lambda_0)$ is the dust mass absorption opacity at $\lambda_0$.  
We adopt a value of $\kappa(\lambda_0) = 1.4$ m$^{2}$ kg$^{-1}$ for a corresponding emissivity of $\beta = 2$, from the \cite{Zubko2004} BARE-GR-S model for a combination of PAHs, graphite, and silicates.
This is the same composition used by \citet[][in their ``standard model'']{Galliano2011} and \cite{RR13}, providing a consistent basis for comparison with their works. 
$\beta$ is kept fixed to 2.0 for use with this reference $\kappa$ value, leaving M$_\mathrm{dust}$ and T as two remaining free parameters.

\subsection{Fitting: Parameter and Uncertainty Estimation}
\label{sec:SEDfit}

Parameter estimation is performed here using Maximum Likelihood Estimation (MLE), by maximizing the log-likelihood function (or, equivalently, minimizing its negative).  
The MLE formalism is invoked here instead of the simpler least squares regression so that censored data (upper limits) can be considered in the fits.
This becomes particularly useful in the determination of the fit errors, discussed below, where the fluxes are varied within their uncertainties and potentially fall above the limit levels.
In the case where all fluxes are detections and errors are gaussian, the MLE result reduces to the familiar least-squares value. 

The log-likelihood function is constructed as follows.  Assuming Gaussian-distributed errors, the probability of $n$ detected bands $i$ having their measured values $F_i, F_{i+1}, \ldots F_n$ is
\mbox{$\prod \frac{1}{\sqrt{2\pi}\sigma_i} \mathrm{exp}\left({-\frac{1}{2}\left( \frac{F_i-F_{m,i}(M,\beta,T)}{\sigma_i}\right)^2}\right) $}, where $F_{m,i}(M,\beta,T)$ is the model Equation~\ref{eq:modifiedBB} evaluated at band $i$ for parameters $M, \beta,$ and $T$.  The appropriate log-likelihood function to optimize is then
\begin{equation} \label{eq:loglike_det}
ln \, \mathcal{L}_i(F|M,\beta,T) = -\frac{1}{2} \sum\limits_i \left( \frac{F_i-F_{m,i}(M,\beta,T)}{\sigma_i}\right)^2.
\end{equation}

An additional term can be introduced to account for upper limits.  A non-detection in band $j$ could result from any integrated flux lower than the limit $F_{lim,j}$.  Therefore, the probability of that flux is determined by integrating the probability distribution over values below the limit:
\mbox{$P_j = \int_{-\infty}^{F_{lim,j}} \mathrm{exp}\left[-\frac{1}{2}\left(\frac{F_j-F_{m,j}(M,\beta,T)}{\sigma_j}\right)^2\right] \, dF_j $}.  Utilizing the Gauss error function, the log-likelihood for the limits can be written as
\begin{multline} \label{eq:loglike_lim}
ln \, \mathcal{L}_j(F|M,\beta,T) = \\  \sum\limits_j \, ln \left( \sqrt{2\pi} \sigma_j \left[ 1 + \mathrm{erf}\left( \frac{F_{lim,j}-F_{m,j}(M,\beta,T)}{\sqrt{2}\sigma_j} \right) \right]\right).
\end{multline}
A similar derivation of an upper limit term for MLE is described by \cite{Sawicki2012}. The total log-likelihood function can then be expressed by combining those of the detections and limits:
\begin{equation} \label{eq:loglike_tot}
ln \, \mathcal{L}_{tot} = ln \, \mathcal{L}_{i} + ln \, \mathcal{L}_{j} ,
\end{equation}
with $ln \, \mathcal{L}_{i}$ reducing to the standard $\chi^2$ in the case where all bands are detected.

Uncertainties on the fit parameters are determined by bootstrap resampling -- generating a distribution of additional fits after resampling the flux densities within their error bars.  
The uncertainties we quote are the differences between the MLE best fits and the 16$^\mathrm{th}$, and 84$^\mathrm{th}$ percentiles of the distributions from 10,000 bootstrap samples. 
An additional source of uncertainty that is not formally considered in the fits is the assumed value of $\kappa$, which can potentially vary substantially depending on the specific variety of dust considered, and could modify our inferred $M_\mathrm{dust}$ by factors of $\sim$ a few.  
See, e.g., \cite{Clark2015}, \cite{Whitworth2019}, and \cite{Cigan2019} for summary discussions of the variety of $\kappa$ values used in the FIR.  

Color corrections to the flux densities, which are typically on the order of a few percent or less and dependent on the shape of the SED (specifically, $\beta$ and temperature), were determined and applied as part of the fitting process.  
First, a preliminary fit to the data was performed to determine the color correction values at each \herschel\ band.  
Then a second fit was performed on the color-corrected data to obtain the true parameters.  
As the temperature value can vary slightly from the preliminary to the second fit, the color correction corresponding to the true fit can also differ slightly, though the flux density difference generally much smaller than 1\%. 
The 100, 160, 250, 350, and 500\um\ color corrections determined for each aperture are given in the final column of Table~\ref{table:SEDs}.

The SEDs and fits are shown in Figures~\ref{fig:SEDfits} (full system apertures) and~\ref{fig:SEDfits_miniregs} (the smaller localized apertures).  
Parameters derived from the model fits are summarized in Table~\ref{table:SEDs}.  Statistics including the minimum, median, and maximum fit values are included for the \lt\ derived in this work, alongside the DGS and KINGFISH samples for comparison.  
The values of $T$, $M_\mathrm{d}$, and $\beta$ for the DGS and KINGFISH galaxies were taken from the modified blackbody fits of \cite{RR13}. 

The number of components in the modified blackbody model can affect the parameter fits, and allowing for a second component \citep[see, e.g.,][]{Dunne2001,Clark2015} can improve the results; a single component fit  will blend aspects of any dust populations contained within the measurements.  For example, if the dust is primarily populated by two differentiable temperatures, then the overall SED will be broader than for a single component, which might correct for this by giving a higher temperature paired with a lower mass.  Single-component fitting can also lead to the appearance of a sub-mm excess, which could be accounted for with a second (cold) component \citep[e.g.,][]{Clark2015}.  
In the following analysis, a single-component model is used for DDO 69, DDO 75, and DDO 210, and results in good fits to the data with small residuals.  
The FIR emission profiles for DDO 70 however (for the full and reduced apertures) are quite broad, and single component fits with $\beta$=2.0 result in notable residual emission with the shape of a blackbody. 
Therefore, for DDO 70 we employ a second component to our fits to better characterize the observed emission. 
Color corrections were not applied for the DDO 70 two-temperature fits. 
They are typically small for the other sources in this sample though, $\sim$3\% at 100\um\ and $\sim$1\% at 500\um, and have a much smaller potential impact on derived masses than the assumed dust model and the considerations discussed in the following section.

\begin{figure*}[ht!]
\centering
	\begin{subfigure}[]{}
		\includegraphics[width=0.48\linewidth]{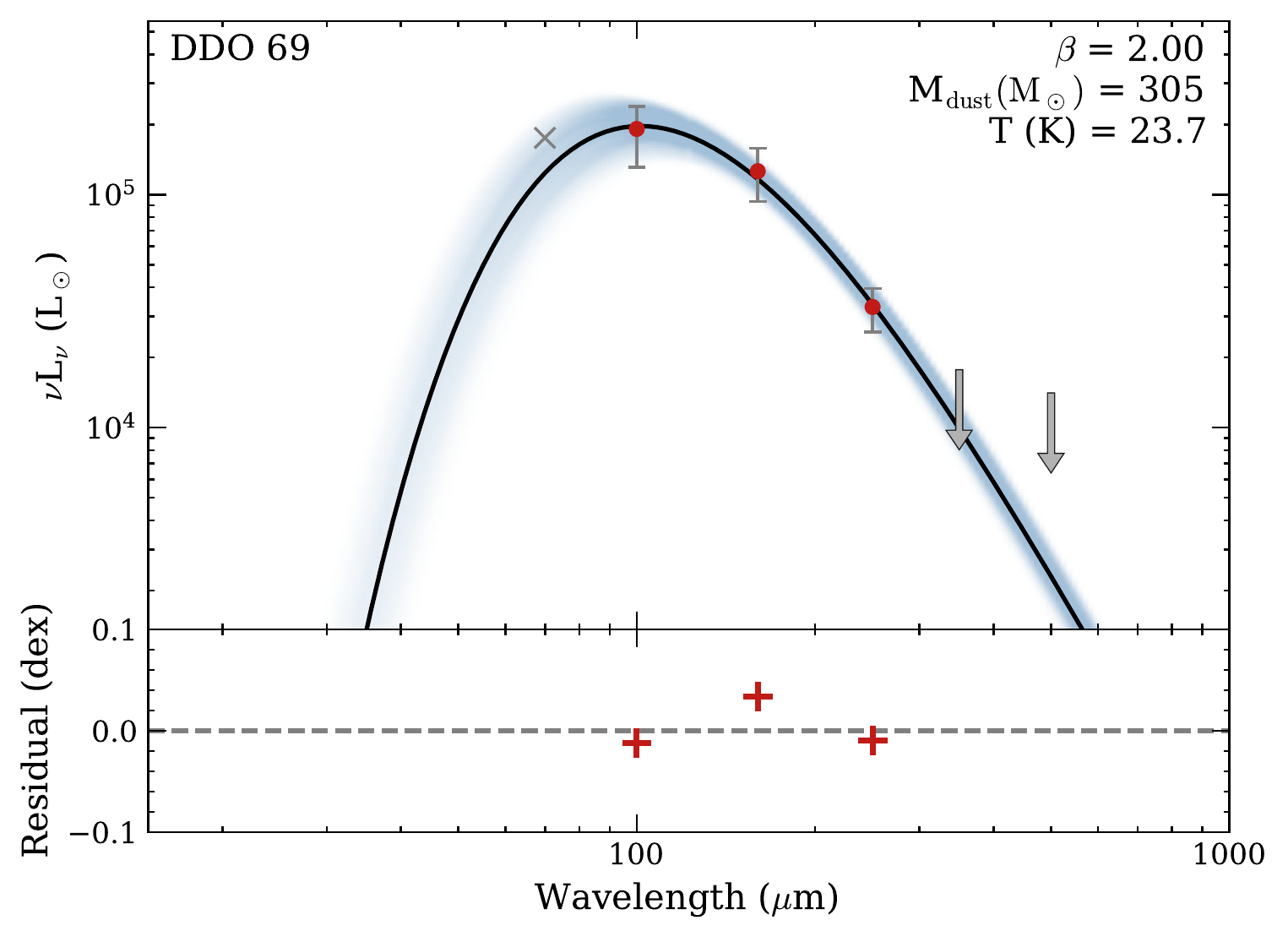}
	\end{subfigure}
	\begin{subfigure}[]{}
		\includegraphics[width=0.48\linewidth]{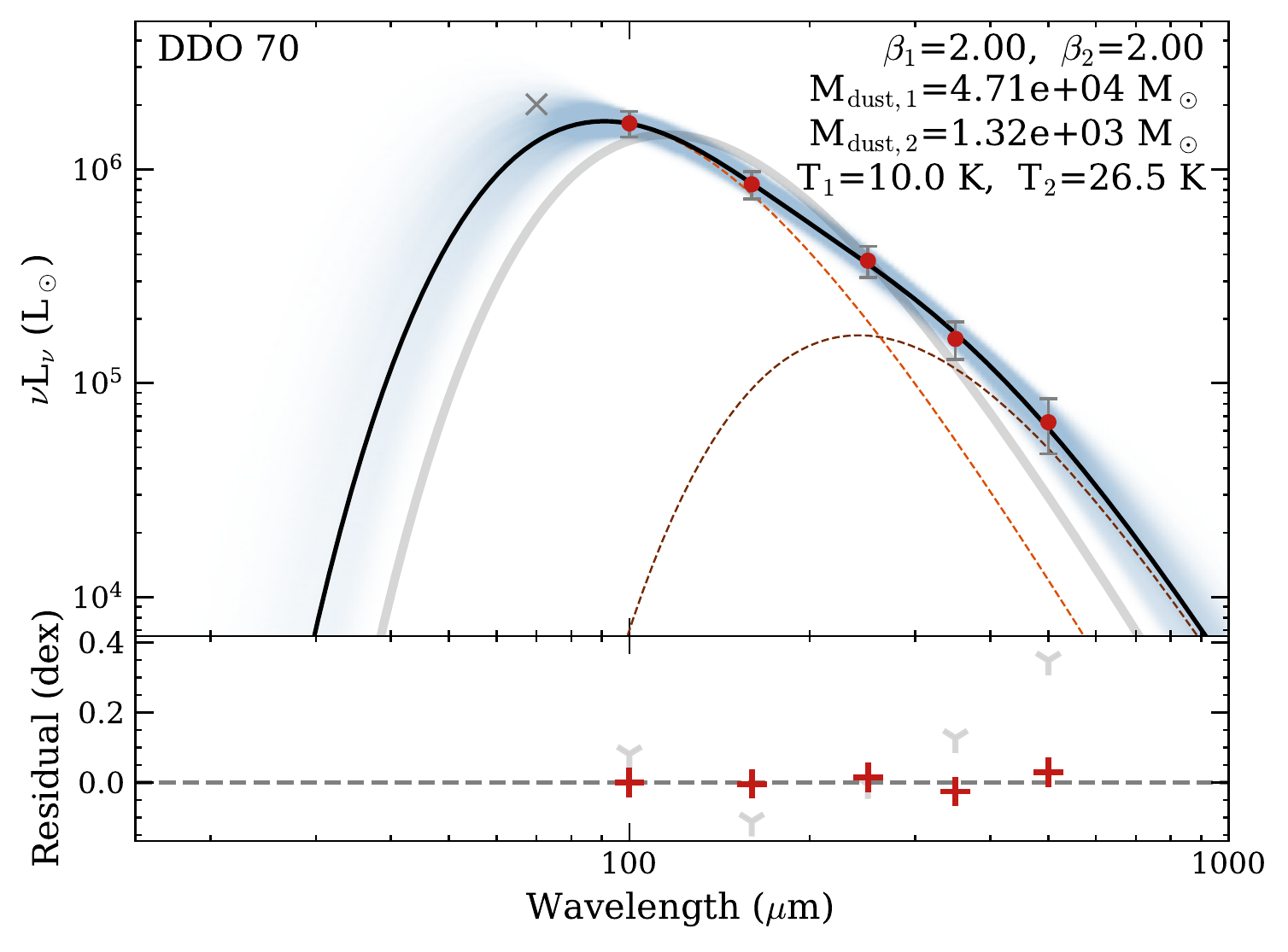}
	\end{subfigure}
	\\
	\begin{subfigure}[]{}
		\includegraphics[width=0.48\linewidth]{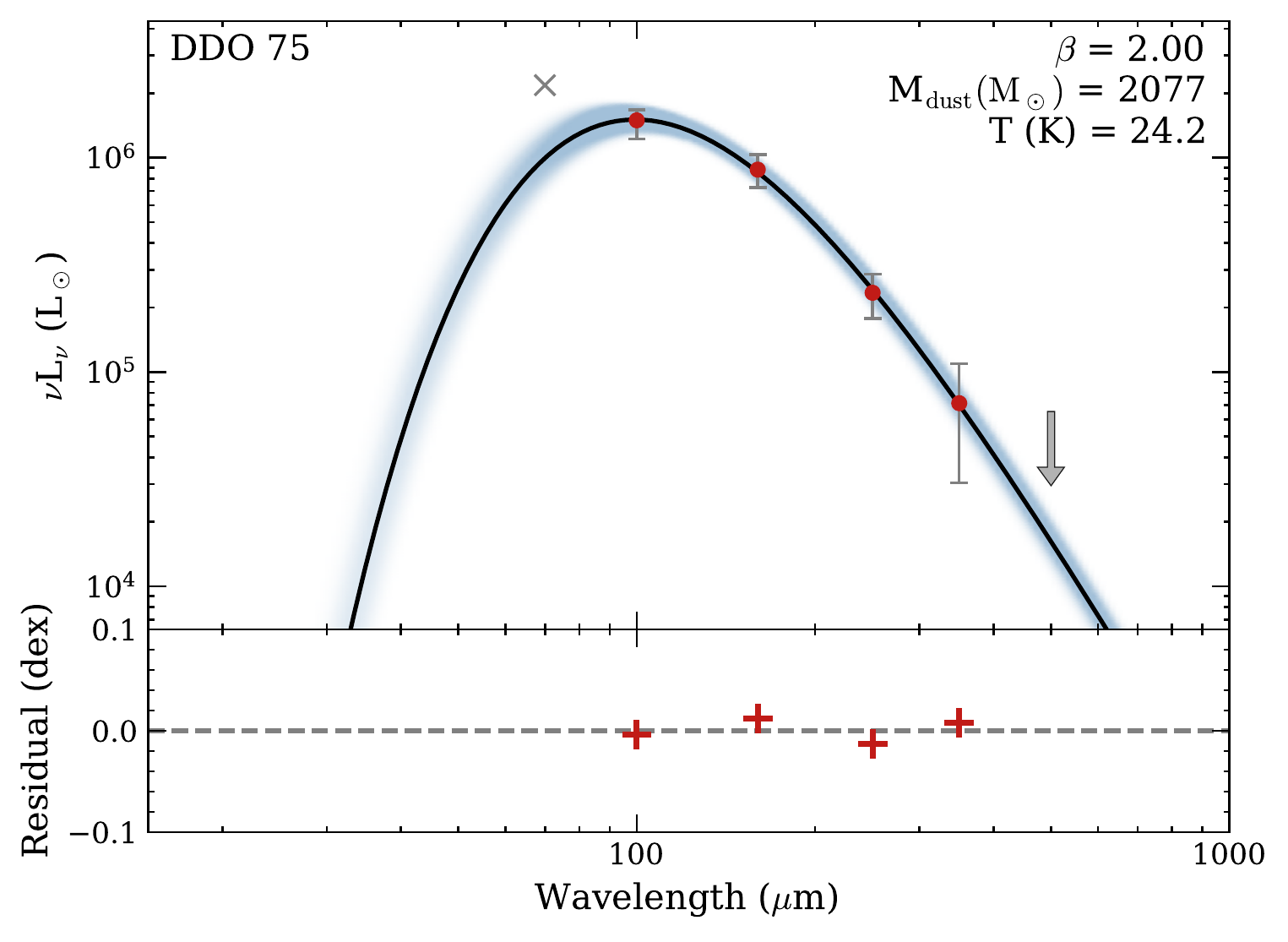}
	\end{subfigure}
	\begin{subfigure}[]{}
		\includegraphics[width=0.48\linewidth]{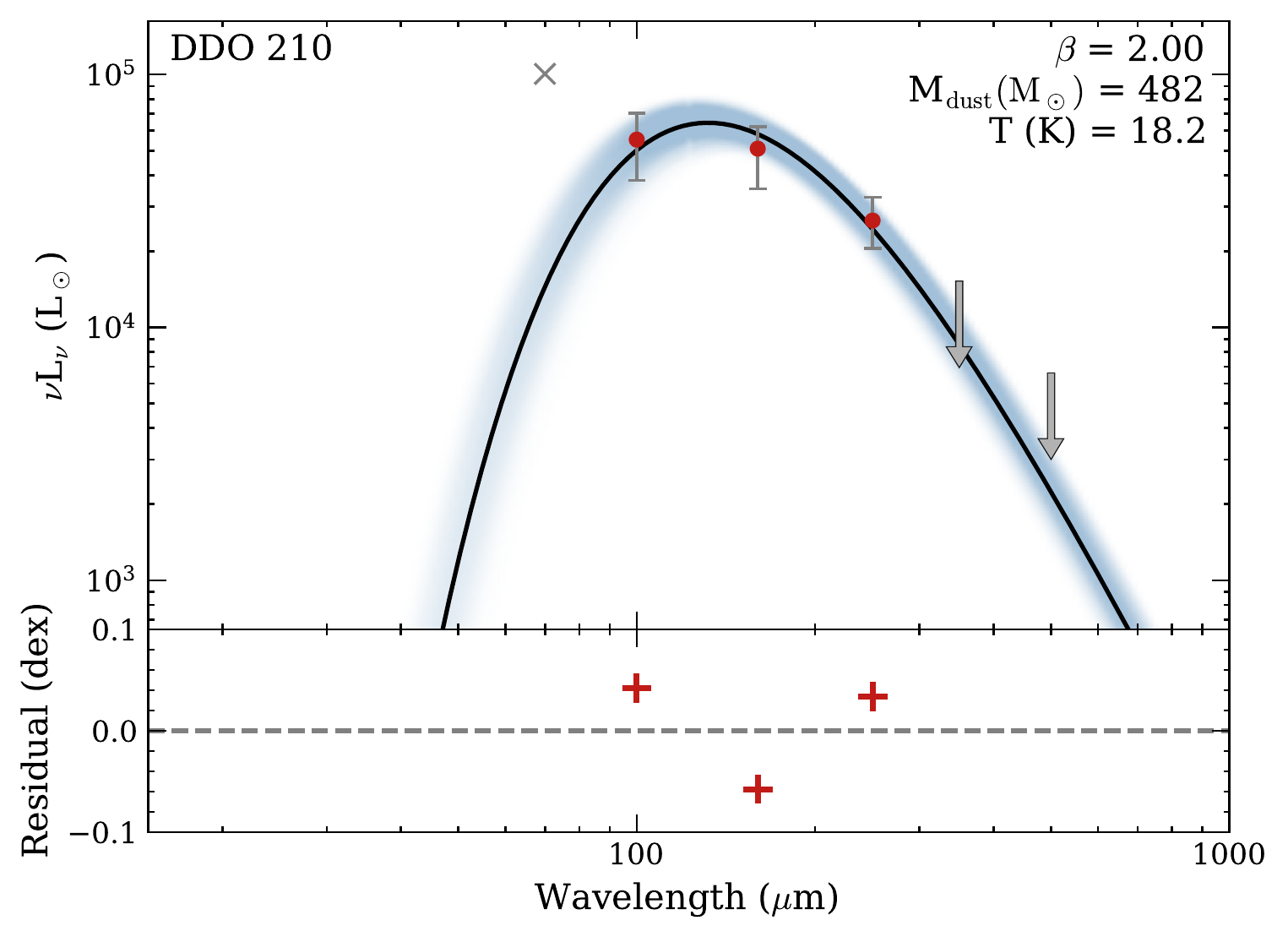}
	\end{subfigure}
\caption{Maximum likelihood fits of the flux densities to a modified blackbody function. The red points are the observed data.  
Upper limits are denoted as gray arrows.  
The blue shading corresponds to the density of bootstrap resample fits, and reflects the uncertainty in the overall fit. 
The MIPS 70\um\ values are shown for reference as grey crosses. 
The lower sub-axes indicate the residual differences between the data and the best fit. 
Both components are shown in color with the combined profile for DDO 70, and the single-component fit plus residuals for this target are shown in gray to illustrate the improvement from including a second component. 
The observed data points include color corrections, except for DDO 70. 
\\ }
\label{fig:SEDfits}
\end{figure*}

\begin{figure}[]
\centering
\includegraphics[width=1.\linewidth]{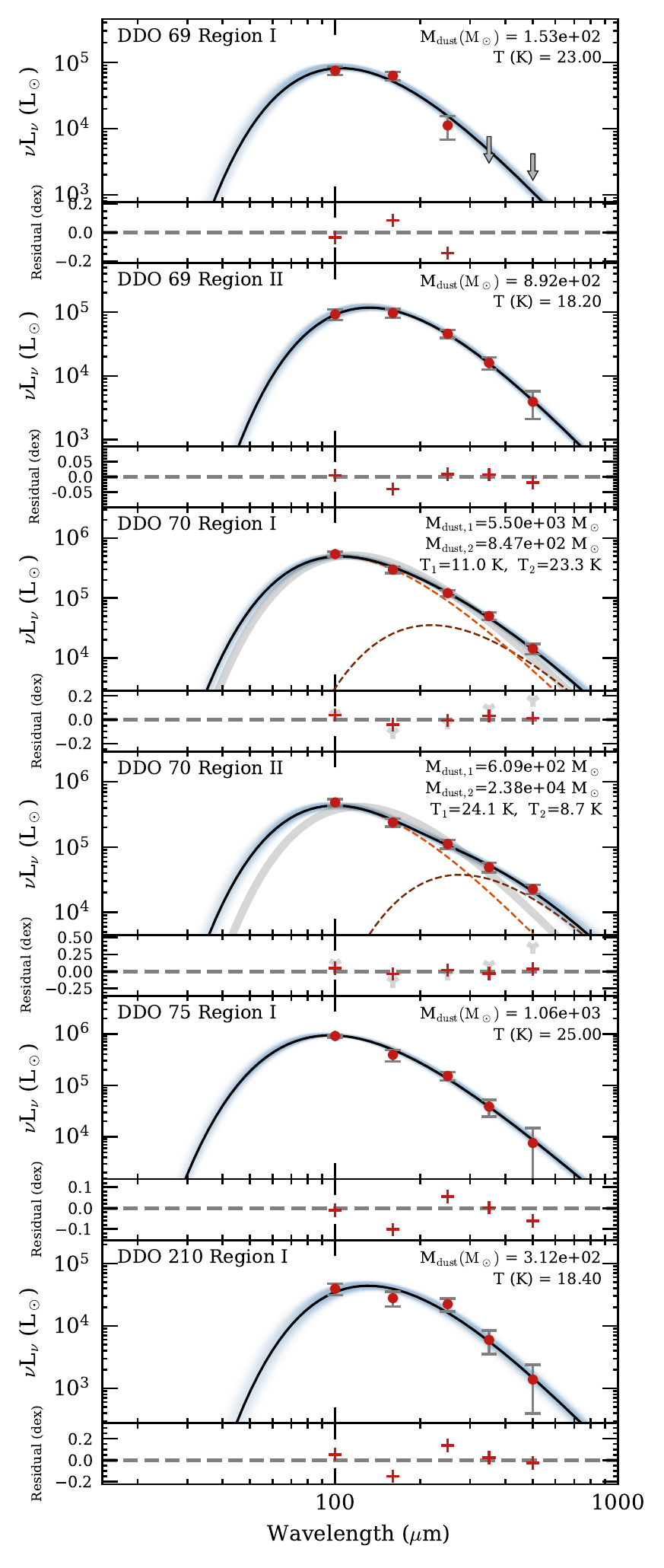}
\caption{Modified blackbody fits for the small individual regions within each galaxy. The red points are the observed data.  Upper limits are denoted as gray arrows.  The blue shading corresponds to the density of bootstrap resample fits, and reflects the uncertainty in the overall fit. The sub-axis below each fit profile indicates the residual differences between the data and the best fit.
For DDO 70, the individual components are shown in color with the combined profile, and its single-component fits plus residuals are shown in gray for reference, showing clear residual structure. 
The observed data points include color corrections, except for DDO 70. 
}
\label{fig:SEDfits_miniregs}
\end{figure}

\begin{deluxetable*}{ lccchccc }
\tablecaption{ Modified Blackbody Fit Results \label{table:SEDs}} 
\tablehead{ \colhead{Galaxy} & \colhead{Region} & \colhead{Temperature (K)} & \colhead{$M_\mathrm{dust}$ ($M_\odot$)} & \nocolhead{$\beta$} & \colhead{$L_\mathrm{TIR}$ ($L_\odot$)} &  \colhead{TIR/FIR} & \colhead{Color Corrections}  }
\startdata
DDO 69 & Whole & $23.7 ^{ +1.5 } _{ -1.4 }$ & ${ 3.1 } ^{ +0.9 } _{ -0.7 } \,\times { 10^2 }$ & $2.00 ^{ +0.00 } _{ -0.00 }$ & ${ 2.1 } ^{ +0.4 } _{ -0.3 } \,\times { 10^5 }$ & $1.0 ^{ +0.2 } _{ -0.2 }$ & 0.97,1.00,0.99,0.98,0.99 \\ \\  
DDO 70 & Whole & $26.5 ^{ +2.1 } _{ -11.6 }$ & ${ 1.3 } ^{ +0.7 } _{ -0.7 } \,\times { 10^3 }$ & $2.00 ^{ +0.00 } _{ -0.00 }$ & ${ 1.9 } ^{ +1.1 } _{ -0.3 } \,\times { 10^6 }$ & $1.6 ^{ +1.1 } _{ -1.1 }$ & \nodata \\ 
       &       & $10.0 ^{ +1.8 } _{ -3.5 }$ & ${ 4.7 } ^{ +2.8 } _{ -5.6 } \,\times { 10^4 }$ & $2.00 ^{ +0.00 } _{ -0.00 }$  & & & \\  \\
DDO 75 & Whole & $24.2 ^{ +1.1 } _{ -1.0 }$ & ${ 2.1 } ^{ +0.5 } _{ -0.4 } \,\times { 10^3 }$ & $2.00 ^{ +0.00 } _{ -0.00 }$ & ${ 1.6 } ^{ +0.2 } _{ -0.1 } \,\times { 10^6 }$ & $1.0 ^{ +0.1 } _{ -0.1 }$  & 0.97,1.00,0.99,0.98,0.99  \\ \\  
DDO 210 & Whole & $18.2 ^{ +1.1 } _{ -1.1 }$ & ${ 4.8 } ^{ +1.8 } _{ -1.3 } \,\times { 10^2 }$ & $2.00 ^{ +0.00 } _{ -0.00 }$ & ${ 6.8 } ^{ +0.8 } _{ -0.8 } \,\times { 10^4 }$ & $1.0 ^{ +0.2 } _{ -0.2 }$ & 0.98,0.96,1.00,0.99,1.00 \\ \\  
\multicolumn{8}{c}{Fit Results For Individual Regions}  \\  \hline \\
DDO 69  & I & $23.0 ^{ +0.9 } _{ -0.8 }$ & ${ 1.5 } ^{ +0.3 } _{ -0.3 } \,\times { 10^2 }$ & $2.00 ^{ +0.00 } _{ -0.00 }$ & ${ 8.7 } ^{ +0.7 } _{ -0.6 } \,\times { 10^4 }$ & $1.0 ^{ +0.1 } _{ -0.1 }$ & 0.97,0.99,0.99,0.98,0.99 \\ \\ 
 & II & $18.2 ^{ +0.5 } _{ -0.5 }$ & ${ 8.9 } ^{ +1.3 } _{ -1.1 } \,\times { 10^2 }$ & $2.00 ^{ +0.00 } _{ -0.00 }$ & ${ 1.2 } ^{ +0.1 } _{ -0.1 } \,\times { 10^5 }$ & $1.0 ^{ +0.1 } _{ -0.1 }$ & 0.98,0.96,1.00,0.99,1.00 \\ \\ 
DDO 70  & I & $23.3 ^{ +0.7 } _{ -0.7 }$ & ${ 8.5 } ^{ +41.2 } _{ -1.6 } \,\times { 10^2 }$ & $2.00 ^{ +0.00 } _{ -0.00 }$ & ${ 5.5 } ^{ +0.4 } _{ -5.3 } \,\times { 10^5 }$ & $1.9 ^{ +1.8 } _{ -1.8 }$ & \nodata  \\ 
 & & $11.0 ^{ +0.8 } _{ -1.5 }$ & ${ 5.5 } ^{ +0.2 } _{ -1.0 } \,\times { 10^3 }$ & $2.00 ^{ +0.00 } _{ -0.00 }$  & & &  \\  \\ 
 & II & $24.1 ^{ +1.0 } _{ -0.9 }$ & ${ 6.1 } ^{ +1.5 } _{ -1.3 } \,\times { 10^2 }$ & $2.00 ^{ +0.00 } _{ -0.00 }$ & ${ 5.0 } ^{ +0.3 } _{ -0.3 } \,\times { 10^5 }$ & $1.8 ^{ +0.2 } _{ -0.2 }$ & \nodata  \\ 
 & & $8.7 ^{ +1.1 } _{ -1.1 }$ & ${ 2.4 } ^{ +1.8 } _{ -0.9 } \,\times { 10^4 }$ & $2.00 ^{ +0.00 } _{ -0.00 }$  & & &  \\  \\  
DDO 75  & I & $25.0 ^{ +0.8 } _{ -0.7 }$ & ${ 1.1 } ^{ +0.2 } _{ -0.2 } \,\times { 10^3 }$ & $2.00 ^{ +0.00 } _{ -0.00 }$ & ${ 9.9 } ^{ +0.5 } _{ -0.4 } \,\times { 10^5 }$ & $1.0 ^{ +0.1 } _{ -0.1 }$ & 0.97,1.01,0.99,0.98,0.99 \\ \\ 
DDO 210  & I & $18.4 ^{ +0.9 } _{ -0.9 }$ & ${ 3.1 } ^{ +0.9 } _{ -0.7 } \,\times { 10^2 }$ & $2.00 ^{ +0.00 } _{ -0.00 }$ & ${ 4.6 } ^{ +0.4 } _{ -0.4 } \,\times { 10^4 }$ & $1.0 ^{ +0.1 } _{ -0.1 }$ & 0.98,0.96,1.00,0.99,1.00 \\ \\ 
\multicolumn{8}{c}{Statistics for the \lt, DGS, and KINGFISH Samples}  \\  \hline \\
Full Regions & Min$|$Median$|$Max & 10.0 $|$ 23.7 $|$ 26.5  & 3.0\textsc{e}2 $|$ 1.3\textsc{e}3 $|$ 4.7\textsc{e}4  & 2.00 $|$ 2.00 $|$ 2.00  & 6.8\textsc{e}4 $|$ 9.0\textsc{e}5 $|$ 1.9\textsc{e}6  & 1.0 $|$ 1.0 $|$ 1.6 & \nodata  \\  
Reduced Regions & Min$|$Median$|$Max  & 8.7 $|$ 20.7 $|$ 25.0  & 1.5\textsc{e}2 $|$ 8.7\textsc{e}2 $|$ 2.4\textsc{e}4  & 2.00 $|$ 2.00 $|$ 2.00  & 4.6\textsc{e}4 $|$ 3.1\textsc{e}5 $|$ 9.9\textsc{e}5  & 1.0 $|$ 1.0 $|$ 1.9 & \nodata \\
\\
DGS  & Min$|$Median$|$Max  & 21.0 $|$ 32.0 $|$ 98.0  & 1.0\textsc{e}2 $|$ 1.2\textsc{e}5 $|$ 2.5\textsc{e}7  & 0.00 $|$ 1.71 $|$ 2.50  & 1.2\textsc{e}7 $|$ 5.3\textsc{e}8 $|$ 5.3\textsc{e}10  & \nodata & \nodata \\ \\
KINGFISH  &  Min$|$Median$|$Max  & 17.0 $|$ 22.5 $|$ 39.0  & 7.0\textsc{e}2 $|$ 1.0\textsc{e}7 $|$ 1.1\textsc{e}8  & 0.00 $|$ 1.95 $|$ 2.50  & 3.0\textsc{e}6 $|$ 4.3\textsc{e}9 $|$ 6.9\textsc{e}10  & \nodata  & \nodata\\ \\
\enddata

\tablecomments{Fit parameters and uncertainties for the modified blackbody function. Fitting was performed using the Maximum Likelihood Method with a term in the probability function to account for upper limits.  Uncertainties were estimated from the 16$^\mathrm{th}$ and 84$^\mathrm{th}$ percentiles of fits from 10,000 bootstrap samples within the flux density errors.   
$L_\mathrm{TIR}$ is determined by integrating the modified blackbody fit from 3--1000\um, and the FIR luminosity is integrated over 50--650\um.
The infrared luminosities and dust masses of the \lt\ galaxies are generally dominated by one or two bright regions. 
The DGS and KINGFISH values are from the modified blackbody fits of \cite{RR13}.
Color corrections were determined for $\beta=2$ and a temperature from a preliminary fit to the data, as described in \S~\ref{sec:SEDfit}; these values are provided in the last column for the 100, 160, 250, 350, and 500\um\ flux densities. 
The fits for DDO 70 consist of two components.  
Color corrections were not applied for this source, though their effect (typically less than $\sim$3\% for the other targets in this sample) should be much smaller than the uncertainties related to the assumed dust model as outlined in \S~\ref{sec:fitresults}. 
}

\end{deluxetable*}

\section{Dust Properties in context with literature samples}

\subsection{Fit Results}
\label{sec:fitresults}

DDO 70 was fit with two modified blackbody components due to a single component with $\beta$=2 being unable to recover the full emission profile observed. 
The residuals were large and had the distinctive shape of a second modified blackbody profile (see Fig.~\ref{fig:SEDfits}), so using an additional component in the fit model is a natural choice.   
Fitted masses range from $\sim$130$M_\odot$ in the warm components to a few times 10$^4 M_\odot$ in the colder components, for combined dust masses of 4.8$\times 10^4 M_\odot$, 6.4$\times 10^3 M_\odot$, and 2.5$\times 10^4 M_\odot$ in the full aperture and regions I \& II, respectively. 
The temperatures ranged from 23--27K and 9--11K for the warm and cold components, respectively. 
The extremely low fitted temperatures in the cold components are not necessarily physical, however -- this could be an artifact of using a simple two-component model, restricting the assumed dust composition (i.e., $\kappa$, $\beta$), insufficiently capturing the grain size distribution, or a number of other considerations.  
Unfortunately, the full set of dust properties cannot be uniquely determined from the relatively few data points available.
However, a variety of approaches to fitting the dust masses can explore possible variations and evaluate the reliability of our general fit results.

Leaving $\beta$ as a free parameter, which can describe the broadness of the FIR emission at the expense of the reliability of the inferred mass, results in $\beta$ values near 0 for all apertures in DDO 70.  
While $\beta = 0$ can be equivalent to pure blackbody emission, a far simpler physical explanation and more realistic scenario is that the dust has a variety of compositions and temperatures.  
To test the effect of varying the dust composition to other commonly assumed varieties, we performed two-component fits to the DDO 70 full galaxy aperture fluxes using a number of combinations of full $\kappa(\lambda)$ curves from the literature, instead of the simpler power law approximations with slope $\beta$.  
For example, using the \cite{Weingartner2001} LMC model dust with \cite{Draine1984} graphite results in a combined dust mass of 5.9$\times 10^4 M_\odot$.  
A combination of \cite{Zubko2004} ACAR and \cite{Jaeger2003} enstatite dust varieties results in 7.7$\times 10^4 M_\odot$.
Generally, all combinations we tested return a warmer component of $\sim$25--30K and hundreds of $M_\odot$, with a cold component of $\sim$10K and order 10$^3$--10$^4$ solar masses of dust -- very similar to our adopted model. 

We also explored two SED template fitting packages, CIGALE \citep[e.g.,][]{Boquien2019} and MAGPHYS \citep{daCunha2008}, to test the variation of different approaches to fitting dust masses.   
In both cases, after moderate tuning of input parameters, reasonable fits to the overall SED of DDO 70 was achievable, and yielded dust masses of 8--9$\times 10^4 M_\odot$, again similar to our simple modified blackbody fits.
This makes sense, as these models use similar dust compositions to those considered above.  
We note, however, that while the overall SED fits were generally good, the FIR spectrum was less well-fit, appearing quite similar to the single-component $\beta=2$ fit shown in gray in Fig.~\ref{fig:SEDfits}.
Finally, a comparison of our MLE fits with a hierarchical Bayesian fit \citep[see, e.g.][for an extensive discussion on the topic]{Galliano2018} yielded nearly identical results, so we do not expect significant bias from our choice of parameter estimation routine.

Considering all of this, our estimated dust masses are relatively robust to variations in assumed dust composition and a wide variety of fitting methods and models. 
Despite the questionably low temperature of the cold component, the combined mass is still relatively consistent with estimates from other methods.   
The uncertainties reported in Table~\ref{table:SEDs} are appropriate for the individual fits, though based on the different analysis methods above we consider wider ranges of uncertainty in later sections -- for DDO 70 for example, on average this ranges from as low as a few hundred solar masses for the warmer individual components, to $\sim$25\% of the fit masses.

The majority of the fits in this new sample, considering both the full-galaxy and smaller apertures, have temperatures around 20K.   
The \lt\ galaxies fall toward the low end of the range of dust temperatures seen in the DGS galaxies, the bulk of which are between $\sim$20 to $\sim$50 K (with two extreme cases of $\sim$90 K).
This may be expected to first order, since many of the DGS dwarfs are undergoing strong bursts of star formation, and more hot stars will heat the gas and dust to higher temperatures.

The dust mass fits for the \lt\ galaxies are all at the very low end of what is seen in the DGS sample, whose masses range from 100 to $2.5\times10^7$ $M_\odot$ with a median of $1.2\times10^5$ $M_\odot$, while this new sample of dwarfs has dust masses ranging from 305--4.7$\times 10^4$ $M_\odot$ for the full apertures, and 153--2.4$\times 10^4$ $M_\odot$ for the smaller regions.

We see that the infrared maps in our galaxies tends to be dominated by one or two bright regions (c.f. Figs.~\ref{fig:mapcomp} and \ref{fig:mapcomp2}).  
These individual regions can therefore constitute from 20\% to nearly all of the total dust mass (and integrated luminosities) of their host galaxies. 
This could have an impact on studies of distant dwarf galaxies -- since observations would effectively only recover the brightest regions, the bulk properties of the galaxy could be misinterpreted if it is not resolved.
We refer the reader to \cite{Polles2019} for a study of IC 10 based on infrared spectroscopy at different spatial scales.

Models of dust emission \citep[e.g., ][]{Galliano2003} predict that smaller dust molecules, such as PAHs and graphites, are generally warmer than larger grains, such as the mixed amorphous carbons and silicates.  Laboratory measurements of different dust varieties \citep[e.g.,][]{Agladze1996} have found that smaller $\beta$ values ($\sim 1$) are more consistent with smaller carbonaceous grains while larger $\beta$ values ($\sim 2$) are more consistent with physically larger grains such as silicates.  
While we should reasonably expect that the true dust content of these galaxies is a mixture of a various different grain species, sizes, and temperatures, further observations will be necessary to disentangle differences in populations.
However, the simple physical model employed here matches the sparsely sampled observed emission profiles reasonably well, allowing us to characterize the combined bulk dust properties to first order.

\subsection{Scaling relations with Metallicity}

\begin{figure}[]
\centering
\includegraphics[width=1.\linewidth]{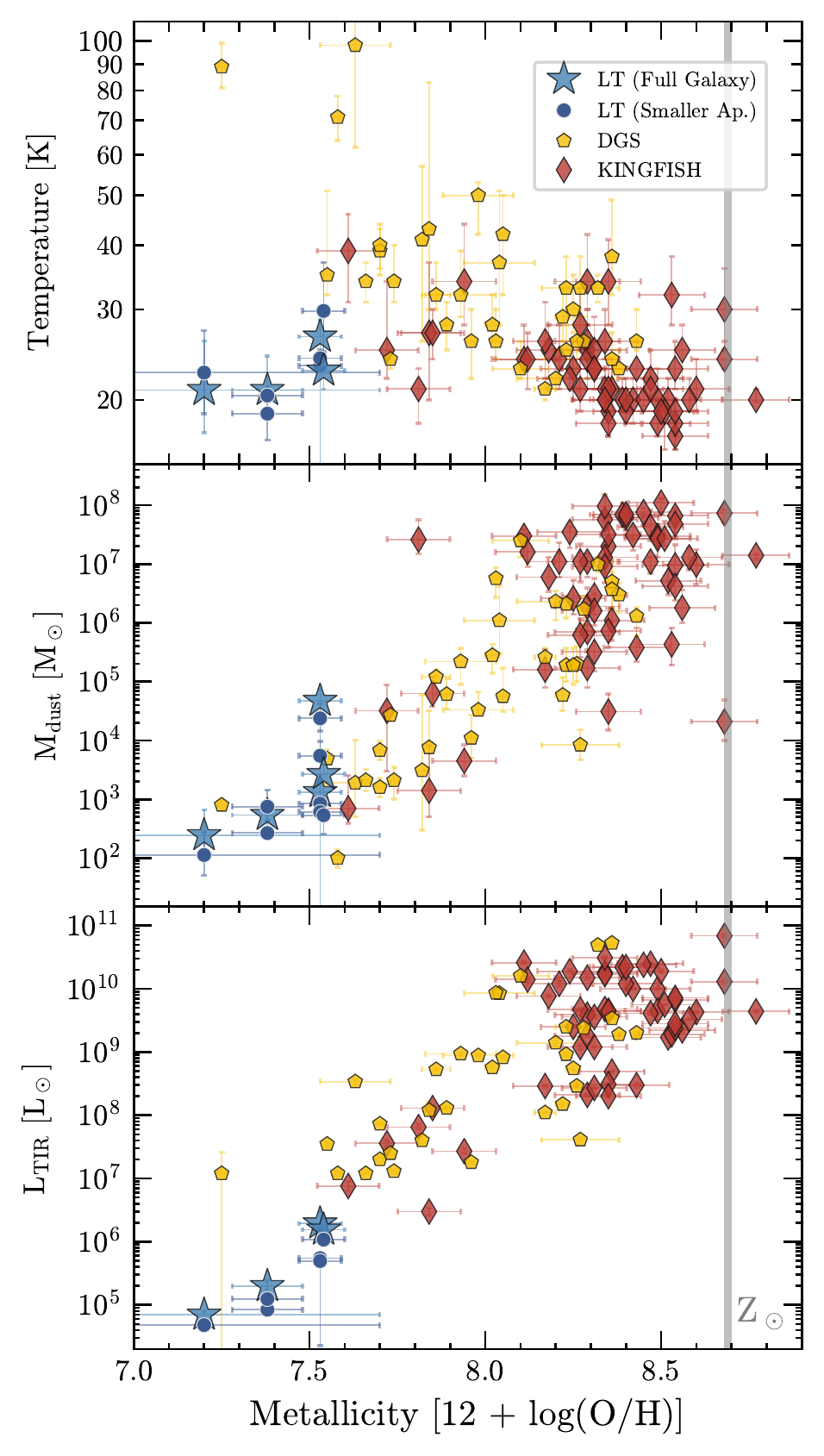}
\caption{Blackbody fit parameters plotted against metallicity for the \lt, DGS, and KINGFISH samples. 
The DGS and KINGFISH values are taken from \citep{RR13}.  
The \lt\ dwarfs observed in this work have mostly cold and cool dust temperatures, and have among the lowest levels of detected $L_\mathrm{TIR}$ and $M_\mathrm{d}$ seen in the DGS sample.
}
\label{fig:parsvsmetal}
\end{figure}

Figure~\ref{fig:parsvsmetal} shows the model fit parameters explicitly compared to metallicity for the \lt, DGS, and KINGFISH samples.  
There are two obvious trends with metallicity: as $Z$ decreases, so does the dust mass and FIR luminosity. 
The $M_\mathrm{d}$ and $L_\mathrm{TIR}$ values in the \lt\ sample follow the metallicity trend tightly, even though the low-$Z$ end of the DGS sample contains some outlying starbursts with high temperatures.

\cite{RR13} noted that the dust masses were two orders of magnitude lower in the DGS galaxies than in the KINGFISH galaxies, but that $M_{\rm dust}$ does not simply scale with galaxy mass.  In particular, they found that as metallicity decreases, the dust mass drops much more rapidly than stellar mass does.  They also determined that it is not an issue with the ``low'' fitted grain emissivities $\beta < 2$ in many DGS galaxies.  Referring to Eq.~\ref{eq:modifiedBB}, a given luminosity that is modeled with a low $\beta$ and a particular mass can alternately be reproduced by increasing $\beta$ (reducing the efficiency) and increasing the mass: that is, by assuming there is more dust, but that it emits less efficiently.  However, when the grain emissivities in the \citet{RemyRuyer2014} sample were fixed to $\beta$=2, the resulting masses were not increased nearly enough to explain the difference.   
Ultimately, \cite{RR13} attribute their difference to the limitations of using a single modified blackbody -- in particular, characterizing only a single grain state and temperature.  
In the \cite{RemyRuyer2015} study, this was shown to be the case when the DGS galaxies were modeled with a more sophisticated SED model covering the mid infrared to sub-mm, instead of a simple modified blackbody.
Still, all the \lt\ dust masses, save for the DDO 70 cold components, are consistent with those derived for one of the lowest metallicity (and notably dust-poor) dwarfs, I Zw 18  \citep{Hunt2014,RemyRuyer2015,Schneider2016}. 

All three samples show a large spread of temperatures near 12+log(O/H)~$\sim$~7.6.  The derived single-component temperatures for the KINGFISH galaxies hint at a slight decrease in $T_\mathrm{dust}$ with increasing metallicity but the DGS and \lt\ samples do not show this.  With many outliers and  large error bars, there is no real trend when all the samples are considered in concert.

All of the \lt\ TIR luminosities, which have a maximum of $1.9\times10^6$ $L_\odot$, are less than the lowest DGS FIR luminosity of $1.2\times10^7$ $L_\odot$.  
Given the small galaxy masses in the \lt\ sample, we would indeed expect them to be less luminous in the FIR than galaxies such as the KINGFISH spirals with dust reservoirs many times larger.  
However, at the very low metallicity end (below $\sim$7.7), the dust masses in all three samples are roughly similar within the errors, while the luminosities appear to decrease more steeply in the new \lt\ sample.

\subsection{Dust-to-Gas Ratios}
\label{sec:dustgasratios}

\begin{deluxetable*}{ l|cccc|cccccc }[]
\tablecaption{Dust-to-Gas Ratios  \label{tab:DGR}}
\tablecolumns{11}
\tablehead{ \multicolumn{1}{c}{ } & \multicolumn{4}{c}{Whole Galaxy} &  \multicolumn{6}{c}{Reduced Apertures} \\ \colhead{} & \colhead{DDO 69} & \colhead{DDO 70} & \colhead{DDO 75} & \colhead{DDO 210} & \colhead{DDO 69 \textsc{i}} & \colhead{DDO 69 \textsc{ii}} & \colhead{DDO 70 \textsc{i}} & \colhead{DDO 70 \textsc{ii}} & \colhead{DDO 75 \textsc{i}} & \colhead{DDO 210 \textsc{i}} }

\startdata
log M$_\mathrm{dust}$ [M$_\odot$]  & 2.48 & 4.68 & 3.32 & 2.68 & 2.18 & 2.95 & 3.80 & 4.39 & 3.03 & 2.49 \\ 
log M$_\mathrm{HI}$   [M$_\odot$]  & 6.69 & 7.08 & 7.19 & 6.08 & 5.97 & 5.62 & 6.26 & 6.31 & 6.60 & 5.65 \\ 
log M$_\mathrm{H_2}$  [M$_\odot$]  & 8.21 & 8.01 & 8.15 & 7.08 & 7.49 & 7.14 & 7.19 & 7.24 & 7.57 & 6.65 \\ 
log M$_\mathrm{gas}$  [M$_\odot$]  & 8.36 & 8.20 & 8.33 & 7.26 & 7.64 & 7.29 & 7.37 & 7.42 & 7.75 & 6.82 \\ 
 \hline 
Dust-HI Ratio   & 6.19$\times 10^{-5}$ & 4.00$\times 10^{-3}$ & 1.35$\times 10^{-4}$ & 3.98$\times 10^{-4}$ & 1.65$\times 10^{-4}$ & 2.13$\times 10^{-3}$ & 3.49$\times 10^{-3}$ & 1.20$\times 10^{-2}$ & 2.66$\times 10^{-4}$ & 7.06$\times 10^{-4}$ \\ 
Dust-Gas Ratio  & 1.33$\times 10^{-6}$ & 3.08$\times 10^{-4}$ & 9.63$\times 10^{-6}$ & 2.64$\times 10^{-5}$ & 3.54$\times 10^{-6}$ & 4.58$\times 10^{-5}$ & 2.69$\times 10^{-4}$ & 9.25$\times 10^{-4}$ & 1.90$\times 10^{-5}$ & 4.68$\times 10^{-5}$ \\ 
\hline 
\enddata
\tablecomments{
Typical uncertainties for the dust masses (see \S~\ref{sec:fitresults}) and HI are of order 25\% and  10\%, respectively, resulting in Dust-HI ratio uncertainties of $\sim$30\%.
The uncertainties in M$_\mathrm{H_2}$ and M$_\mathrm{gas}$ are estimated as $\sim$2--3 times the inferred masses, and therefore the derived Dust-Gas ratios are uncertain by similar factors. 
}
\end{deluxetable*}

\begin{figure*}[]
\centering
\includegraphics[width=1.\linewidth]{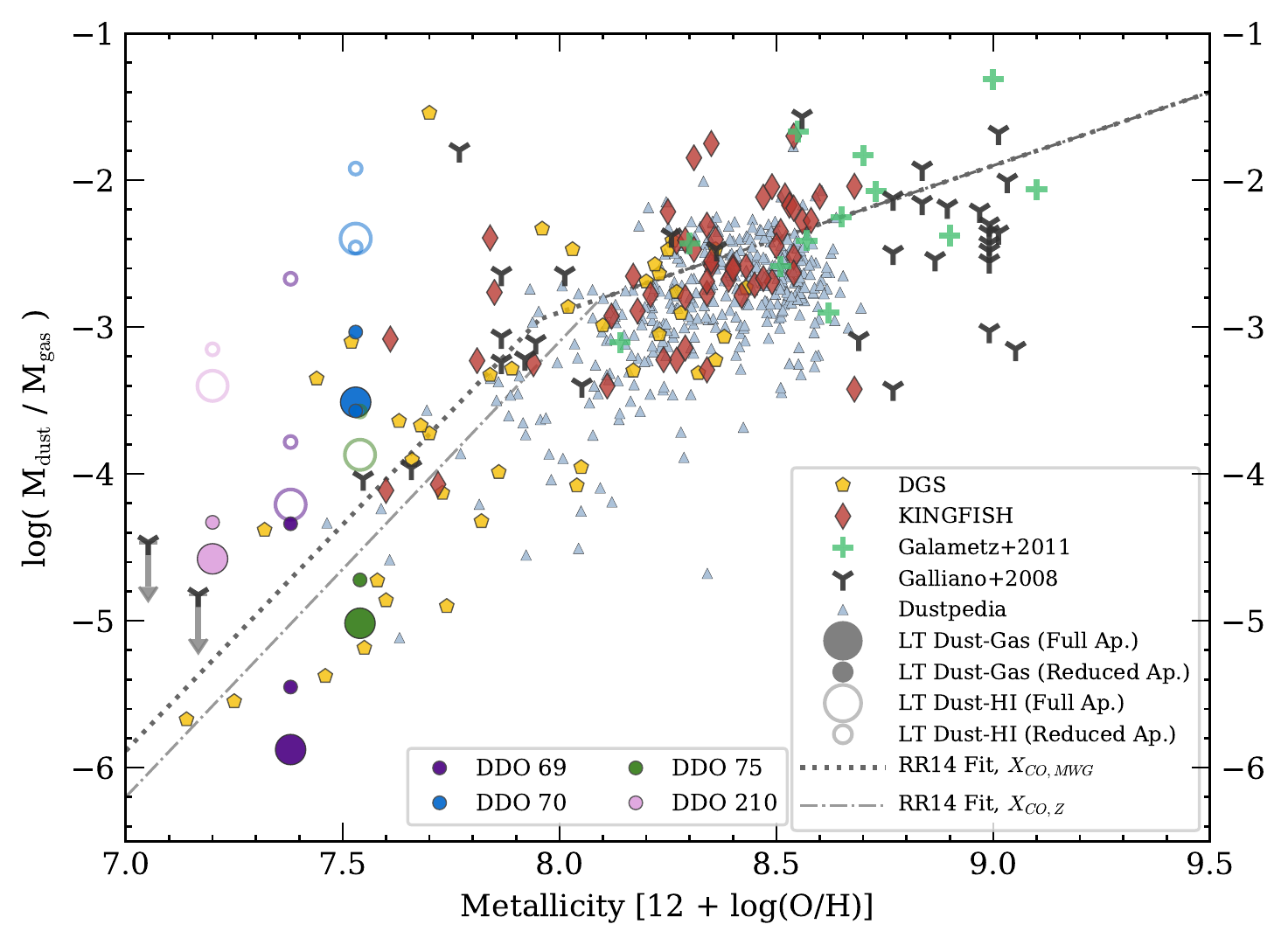}
\caption{Dust to Gas Ratios with metallicity.  
The uncertainties on the ratios are factors of $\sim$2--3 (see text).  
Dust to \hi\ ratios are also shown for the \lt\ sample, with typical uncertainties of order 30\%.. 
We note that the \htwo\ masses, and therefore $M_\mathrm{gas}$, in our galaxies were determined from \cii\ measurements instead of CO which was used in other samples.  The KINGFISH, \cite{Galametz2011}, \cite{Galliano2008}, and Dustpedia \citep{DeVis2019} samples are overplotted for reference.  The DGS and KINGFISH metallicities are taken from \cite{DGS} and \cite{KINGFISH}, respectively.  For a full list of references for the gas masses in those samples, we refer the reader to Table A.1 of \cite{RemyRuyer2014}.  
The dotted and dash-dotted lines are broken power law fits to the DGS data from \cite{RemyRuyer2014} for $X_{\rm CO}$ taken as the Milky Way value, and scaled with $Z$, respectively. The \lt\ sample roughly agrees with their noted trend of decreasing DGR with decreasing metallicity. 
}
\label{fig:DGR}
\end{figure*}

The dust masses derived in \S~\ref{sec:SEDs} can be compared with the exquisite \lt\ \hi\ maps \citep{Hunter2012LTdata} to derive ratios of the dust and gas content in these galaxies.  
The gas mass in a galaxy for this purpose consists primarily of \hi+\htwo, with smaller amounts of helium and gas-phase metals.  
The \hi\ mass can be calculated from the optically thin limit to the radiative transfer equation as
\begin{equation} \label{eq:M_HI}
M_\mathrm{{\rm H{\sc I}}} [M_\odot] = 235.6 \ (D[\mathrm{Mpc}])^2 \ S [\mathrm{Jy \ m/s}]
\end{equation}
using the integrated emission $S$ in the region of interest.  

\htwo\ masses are typically estimated from CO measurements in normal galaxies.  
However, CO is exceedingly difficult to detect in low-metallicity systems \citep[][for example]{Leroy2009,Cormier2014}. 
We follow a method of relating \cii\ to hydrogen in a similar manner to that of \cite{Madden1997}, who determined the amount of \htwo\ for the low-$Z$ dwarf IC 10, and found it to be a useful method when CO is not a suitable tracer. 
We also refer the reader to more recent works using $I_\mathrm{[C\textsc{II}]}$ instead of CO as a proxy for \htwo, by \cite{Velusamy2010} and \cite{Langer2014} for the Milky Way, as well as \cite{Jameson2018} for the Magellanic Clouds, as additional examples.

The \htwo\ masses of these \lt\ galaxies are estimated from the \cii\ measurements of \cite{Cigan2016}, who determined that most of the observed \cii\ in these systems originates from photodissociation regions as opposed to the ionized ISM.
\cite{Polles2019} found the same for IC 10, and \cite{Cormier2019} have shown that the fraction of \cii\ corresponding to ionized gas is expected to decrease with metallicity, down to at most a few percent at the metallicities of these galaxies.
For solar-metallicity photodissociation regions (PDRs), \cii\ only originates from a thin outer layer.  
As metallicity decreases, according to the model of \cite{Bolatto1999} for example, the filling factor of the CO-emitting core shrinks compared to that of the \cii-emitting region.
\htwo\ can still be present in the absence of CO -- at low extinction $A_V$ \citep[e.g.,][]{Tielens1985,Madden1997} -- and at the extremely low metallicities of the galaxies considered in this work, \cii\ could trace nearly all of the \htwo\ in PDRs. 
By modelling the gas collisional excitation parameters, the total hydrogen content of the neutral regions can be determined by relating it to the carbon content. 
The molecular hydrogen component is then inferred as the excess beyond the atomic component observed in \hi.

In this prescription, the carbon column density is calculated from \cii\ emission using the optically thin radiative transfer equation (c.f \citealp{Madden1997} Eq.~1, for example): 
\begin{equation} \label{eq:I_Cplus}
I_\mathrm{[CII]} = \frac{h \nu A_{ul}}{4\pi} \left[ \frac{ g_u / g_l \, e^{-h\nu /kT} }{ 1 + g_u / g_l \,	 e^{-h\nu /kT} + n_{crit}/n }  \right] N_\mathrm{C^+} \quad ,
\end{equation}
where $A_{ul}$ is the Einstein emission coefficient, $g_u$ and $g_l$ are the statistical weights of the levels, and $n_{crit}$ is the critical density. 
We model the collision parameters $n$ and $T$ using the PDR Toolbox \citep{Pound2008} with measurements of \cii\ and \oi\ from \cite{Cigan2016}, along with the integrated TIR continuum.   
$N_\mathrm{C^+}$ can be related to the column density of the colliding hydrogenic nuclei $N_\mathrm{H}$, using the local carbon abundance C/H as the conversion factor $X_\mathrm{C^+} = N_\mathrm{C^+} / N_\mathrm{H}$ \citep[e.g.,][]{Madden1997,Langer2014}. 
Instead of scaling the Galactic ISM carbon abundance of 1.42$\times 10^{-4}$ \citep[][from \cii /\hi, assuming equivalence to C/H]{Sofia1997} by metallicity as in other works, we can calculate $X_\mathrm{C^+}$ directly from our \cii\ and \hi\ maps, though both methods give similar results. 
Finally, the neutral gas component measured from \hi\ can be subtracted, leaving the molecular component: $N_{\rm H_2} = (N_{\rm H} - N_{\rm HI})/2$.

The \htwo\ mass is generally several times the \hi\ mass in each galaxy: \cite{Stacey1991} found that
$N_{\rm H_2}/N_\mathrm{H{\sc I}} \sim 5$ is common in many luminous galaxies, and \cite{Madden1997} have argued that the same column density ratio is reasonable for IC 10.  
We find that this is also a reasonable approximation for our galaxies, with calculated $N_{\rm H_2}/N_\mathrm{H{\sc I}}$ values of $\sim$4 for DDO 70 and DDO 75. 
DDO 69 has a higher inferred ratio of $\sim$16.
DDO 210 had no \cii\ measurements, so we adopt $N_{\rm H_2}/N_\mathrm{H{\sc I}} = 5$ for this galaxy.

The total gas mass is calculated as the sum of \hi, \htwo, He, and metals in gas form: $M_{\rm gas}$ = $M_{\rm H{\sc I}}$ + $M_{\rm H_2}$ + $M_{\rm He}$ + $Z_{\rm gal}\times M_{\rm gas}$. If the helium mass is assumed to follow $M_{\rm He}$ = $Y_\odot$ $M_{\rm gas}$, where $Y_\odot$ is the Galactic helium mass fraction of 0.27 \citep{Asplund2009}, then the total gas mass can be written as:
\begin{equation}
M_{\rm gas} = \frac{M_{\rm H{\sc I}} + M_{\rm H_2}}{1-Y_\odot-Z_{\rm gal}} \quad .
\end{equation}
The computed dust-to-gas ratios for the \lt\ sample are reported in Table~\ref{tab:DGR}.
\cite{Jameson2018} quote a factor of two uncertainty on their \htwo\ masses, dominated by the differences from the various models they considered; we favor a more conservative factor of $\sim$2--3 estimate for our galaxies with \cii\ measurements, primarily from the assumptions about how carbon relates to hydrogen in these systems.  
The uncertainty for DDO 210 is higher, as the \htwo/\hi\ ratio could vary from the assumed value by a factor of a few.

Our DGR values are generally very small compared to galaxies with higher metallicity, as expected.  
Though, when plotted (Figure~\ref{fig:DGR}) against DGR values from the DGS and KINGFISH samples \citep[][\footnote{These DGR values were calculated with the updated dust masses from \cite{RemyRuyer2015}, from full SED modeling.  While the $M_\mathrm{d}$ values from \cite{RR13} were preferable for direct comparison to modified blackbody fits as in e.g. Fig.~\ref{fig:parsvsmetal}, the differences in dust masses do not change the overall trends, so we use these DGR values for consistency with the published results.} with $M_\mathrm{gas}$ determined from CO]{RemyRuyer2014}, they broadly fall in line with the trend fit to the DGS galaxies by \cite{RemyRuyer2014}.   
In that study they fit their data with a broken power law, once for CO-based \htwo\ values using the local Milky Way $X_\mathrm{CO}$ factor, and a second time using $X_{\mathrm{CO},Z}$ that is scaled with metallicity.  
The breaks occur near 12+log(O/H)=8 as a result of their findings that $M_{\rm H_2}/M_{\rm H{\sc I}}$ ratios in their sample exhibit a break in that vicinity, as well as the fact that several other studies \citep[such as][]{Draine2007,Galliano2008} have shown that DGR decreases with metallicity with a slope of --1 down to 12+log(O/H) of 8.0--8.2.  
It should be noted that the \htwo\ masses \cite{RemyRuyer2014} infer for metallicities below $\sim$7.8 come not from direct CO detections in those galaxies, but from applying the mean $M_{\rm H_2}/M_{\rm H{\sc I}}$ ratio for CO detected below 12+log(O/H)$\lesssim 8.1$ to the \hi\ mass in each galaxy.
The low-$Z$ DGS dust/gas ratios did not include a CO-dark \htwo\ component, which would steepen the non-linear trend further.

We also include in our comparison the Dustpedia archive \citep{Davies2017}, which gives multiwavelength photometry for 875 local galaxies.  
\cite{DeVis2019} first compared the DGR to $Z$ for this sample, which are included in Fig.~\ref{fig:DGR}.  
The Dustpedia sample fills in more of the parameter space at moderately low metallicities, and complements the picture of a steeper slope below 12+log(O/H)$\sim$8 seen in the other samples, namely the \lt\ DGS dwarfs. 
The dust masses used in these dust to gas ratios for the DGS, KINGFISH, Dustpedia, \cite{Galliano2008}, and \cite{Galametz2011} samples are all derived from SED models rather than modified blackbody fitting, and incorporate different dust models, however the overall trends will be similar despite these differences. 
The dust masses determined from modified blackbodies could be lower than those for the SED models by factors of $\sim$2 to at most 10 \citep[e.g.,][]{RemyRuyer2015}. 

DDO 69 is a particularly interesting case, in that is has an extremely low DGR for its metallicity -- similar to previous results for I Zw 18 and SBS 0335-052 (as shown by \citealp{RemyRuyer2014} and further discussed by \citealp{Schneider2016}) and a few sources in the H{\sc I}-rich sample from \citet{DeVis2017a,DeVis2017b}, suggesting some dwarf galaxies are extremely dust poor at a given $Z$ and gas mass compared to the bulk of dwarfs studied in the literature.
Properly accounting for a CO-dark \htwo\ component in the lowest-metallicity DGS galaxies could push them even lower than their extreme positions in Fig.~\ref{fig:DGR}.
Even ignoring the \htwo\ content, the Dust/\hi\ ratios for our galaxies still generally fall below the shallower linear trend line of the higher-metallicity galaxies.

\subsection{Comparison with dust models}

\begin{deluxetable*}{l|lccc}[h!]
  \tabletypesize{\scriptsize}

\tablecaption{Model Summary. \label{tab:models}}
\tablehead{ \colhead{Source} & \colhead{Name} & \colhead{SFH} & \colhead{Dust Source} & \colhead{Other Parameters}  }
\startdata
Z14 & Models 1-3 & Bursty & stardust $+$ grain growth & inflows$+$outflows  \\
 & Model 4 & Continuous & stardust $+$ grain growth & inflows$+$outflows  \\
  & Models 5-7 & Bursty & stardust $+$ grain growth & inflows$+$outflows  \\
& & & \\
DV17b  & Model I  & MW & stardust & \\
 & Model III  & MW  & stardust & outflows, gas mass $=10^9\,\rm M_{\odot}$ \\
 & Model VI  & Delayed  & stardust ($1/100$) $+$ grain growth & inflows$+$outflows \\ 
 & Model VII  & Bursty  & stardust ($1/12$) $+$ grain growth & inflows$+$outflows \\ 
 \hline
This work  & Model A & Delayed & stardust $+$ grain growth ($\times 10$) & destruction ($\times 100$)\\
& Model B & Delayed & stardust $+$ grain growth ($\times 10$) & gas mass ($1/50$) \\ 
& Model C & Delayed & stardust ($1/10$) $+$ grain growth ($1/800$) & gas mass ($1/100$), destruction ($\times 10$) \\ 
\enddata
\tablecomments{Comparison of literature models showing the predicted build up of DGR with metallicity from Z14 - \citet{Zhukovska2014} (all models) and DV17b - \citet{DeVis2017b}. We refer the reader to those works for more details on the models. Here, we only compare Models I, III and VI and VII from \citet{DeVis2017b}.  Models I and III indicate the growth of the DGR with $Z$ for galaxies with efficient stardust production but no grain growth and a Milky Way like SFH \citep{Yin2009}, whereas Models VI and VII include grain growth as an important dust contributor and were shown to be more representative of star forming (Model VII) and more quiescent dwarf systems (Model VI) when assuming a bursty and non-bursty SFH respectively. The \cite{DeVis2017b} models assume an initial gas mass of $10^9\,\rm M_{\odot}$ and Models VI and VII require a reduced fraction of dust formed in stars (shown by the factors in parentheses).  This work takes the \citet{DeVis2017b} model VI but changes input parameters including initial gas mass, dust destruction, efficiency of dust production in stars and in grain growth) to explore lower DGRs with metallicity. This Work Models A-C are therefore the same as DV17b Model VI unless stated otherwise (in parentheseses). }

\end{deluxetable*}

\begin{figure*}[h!]
\centering
\includegraphics[width=0.8\linewidth]{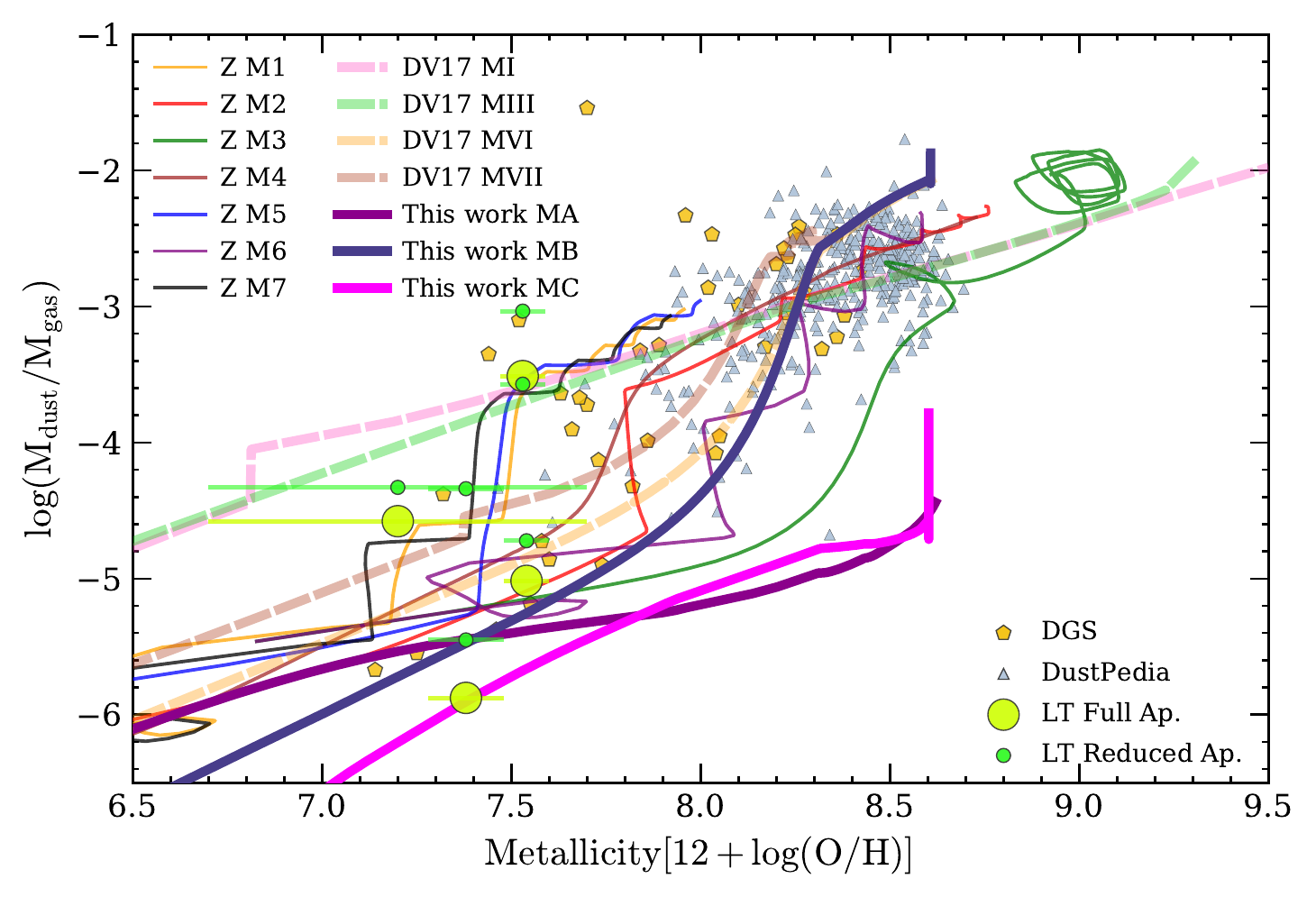}
\caption{Dust/Gas Ratio models for various star formation histories from \citet{Zhukovska2014} (denoted Z M[X] where [X] refers to models 1--7 from their work) and different dust sources from \citet{DeVis2017b} (denoted DV17 M[X] where [X] is one of I, III, VI, or VII corresponding to the models of the same name in their work). For \citet{Zhukovska2014}, the specific evolutionary tracks are determined by the number of bursts, the time and duration, the star formation timescale, and the gas infall timescale. Models range from periodic bursty star formation (step patterns, e.g., Z M1) to continuous (smooth curves, e.g., Z M4).  The \cite{DeVis2017b} models compare different dust sources and include a delayed SFH thought to be appropriate for quiescent dwarfs. We add new models (MA, MB and MC) here by taking the DV17b Model VI (appropriate for gas rich, low star-forming dwarfs) and changing the input parameters to explore lower DGRs than reached in \cite{Zhukovska2014} and \cite{DeVis2017b}.  See Table~\ref{tab:models} for more details.
}
\label{fig:DGRmodels}
\end{figure*}

\cite{Zhukovska2014} recently modeled the evolution of dust and metallicity in dwarf irregulars over time, for a variety of star formation histories (SFH).  Their open-box chemical evolution model allows for the slow infall of primordial gas from outside the galaxy as well as gas outflow from galactic winds, though they calculate that galactic outflows are negligible.  Starting from zero mass, the galaxy forms by accreting infalling gas until reaching the final $M_\mathrm{tot}$ after the specified infall timescale. Dust is formed in these models by stars (AGB and type-II supernovae), and they also consider growth of dust mass by accretion in the ISM onto existing dust grains.  

\citet{Feldmann2015} used similar chemical evolution models to argue that dwarf galaxy samples such as the DGS required strong inflows and outflows to explain their observed properties, though \citet{RemyRuyer2014} showed that the models of \citet{Asano2013} can match the DGS values over a wide range of $Z$ by varying the star formation history. Later, \citet{DeVis2017b} showed that the \citet{Feldmann2015} and \citet{Zhukovska2014} models fit the more actively star-forming dwarfs in the DGS but could not explain more quiescent dwarfs with lower DGRs (e.g., \citealp{RemyRuyer2014}).  
Instead, \citet{DeVis2017b} argued that the gas-rich, less actively star forming dwarfs required different dust sources -- including interstellar grain growth, and reducing AGB stars and supernova dust production rates -- as well as a delayed SFH with less extreme outflows.  We note that \cite{Schneider2016} explained the order of magnitude variation in DGR measured for two dwarfs at the same metallicity as being due to more efficient in-situ dust growth in the ISM in one of the sources where higher cold gas densities were observed.   In Figure~\ref{fig:DGRmodels}, we compare the sample of dwarfs observed in this work with all the models from \citet{Zhukovska2014} and a selection of models from \citet{DeVis2017b} with details in Table~\ref{tab:models}.

When compared to the example evolutionary models of \citet{Zhukovska2014} (labeled Z in Figure~\ref{fig:DGRmodels}), the galaxies align with different star formation paths.  Thus the star formation histories of each galaxy in our sample could be quite different, with a mix of discrete bursts and low-level continuous production.   DDO 210 corresponds most closely to the sample evolutionary tracks with an early burst of SF.  DDO 75 is more closely aligned with the models with smoother early SF activity and some bursts at later times.  DDO 69 falls below all of the models considered by \cite{Zhukovska2014}.

In comparison to the \citet{DeVis2017b} predictions, Models VI and VII reach similar DGRs at the same $Z$ as galaxies DDO 210 and DDO75, but again lie well above the source DDO 69.  We therefore ran the chemical evolution code from \citet{DeVis2017b}\footnote{available at {\url{ https://github.com/zemogle/chemevol/releases/tag/v_de_vis2017}} version v$\_$de$\_$vis2017}
 using their Model VI and changed the input parameters in order to attempt to reach lower DGRs. 
Models A--C (Table~\ref{tab:models} and Figure~\ref{fig:DGRmodels}) respectively show the outcomes of this analysis: A -- the effect of faster grain growth timescales and faster dust destruction rates; B -- decrease in initial gas mass, and faster grain growth; and C -- decrease in initial gas mass, decrease in stardust contributed by supernovae\footnote{In practice reducing this parameter makes very little difference to the predicted DGR with $Z$ compared to the \citet{DeVis2017b} Model VI since their model had already reduced the contribution from Type II SN to the dust budget by $100\times$ compared to typical values used, e.g., \citep{Dwek1998,Morgan2003,Rowlands2014,Zhukovska2014}.}, slower grain growth timescales and faster destruction rates.  
In models VI and A--C, the delayed SFH evolves as $\mathrm{SFH}(t) \propto \tau^{-2} exp(-t/\tau)$ , where $t$ is the age of the galaxy and the SF timescale $\tau$ is 15~Gyr \citep[see][for more details]{DeVis2017b}.  
 
Model A starts with the same DGR at $12+{\rm log (O/H})<6.75$ as Model VI but diverges significantly once metallicities of $12+{\rm log (O/H})>7.25$ are reached. Model B predicts lower values of DGRs by $\sim 0.5$ dex at low metallicities but rapidly increases to match the DGRs reached in Model VI at metallicities of $12+{\rm log (O/H})\sim 8$.  The reduced DGR initially seen in this model due to the reduction in initial gas mass is counteracted at later metallicities when the interstellar grain growth ``kicks in'' (note that this parameter is 10$\times$ faster than Model VI).  Model C lies below the other models by almost a decade in DGR even at solar metallicities and above, due to combination of reduced initial gas mass, slower grain growth timescales, and increased dust destruction rates.

These parameters have different parts to play in the different metallicity regimes, however. A delayed SFH such as those in Models VI and A--C slows down the build up of dust and metals compared to the other tracks shown.  Next, as we decrease the supernova dust production, the amount of dust made early on in the galaxy evolution (at lower metallicities) decreases.  Similarly, we see the same effect when reducing the initial gas mass in Models B and C compared to Models VI, VII and A.   Changing the amount of dust produced in low-intermediate mass stars only affects the DGR at higher metallicities (given the SFHs used here) since these stars take $> 500$ Myr to reach their dust-production stage and therefore cannot be responsible for low DGR ratios such as DDO 69.   

If one fixes the stardust contribution to the assumed values in \citet{Zhukovska2014} Models 1-7 or \cite{DeVis2017b} Model VI or VII, then changing the grain growth timescale in our models has little effect on the DGR values at low metallicities ($12+{\rm log (O/H})<7.7$) since a critical metallicity is required to be reached before grain growth ``kicks in" and becomes the dominant dust producer \citep{Zhukovska2014,Asano2013}. To reach very low DGRs at metallicities above this, one requires extremely long grain growth timescales ($\sim$1000-8000$\times$ slower grain growth rates than typical values used in \citet{Zhukovska2014,Feldmann2015} and \citet{DeVis2017b}). Increasing the dust destruction rates again acts to reduce the DGR, though again the increase needs to be one to two orders of magnitude to move from the parameter space probed by Models  A--C. 

We therefore show that the orders of magnitude variation in DGRs at the same metallicity for chemically young dwarfs might reveal variations in the relative contributions of dust source in these systems (as also seen in \citealt{Schneider2016}). Here we argue that the most dust-poor quiescent dwarfs can be explained by higher destruction rates and extremely slow grain growth timescales compared to more massive systems or more actively star-forming dwarfs.

 An interesting future project would be to compare the star formation histories determined directly from stellar data and color-magnitude diagrams to these models. If measured star formation histories differ significantly from the models, they could provide useful information to revise the dust production models assumed to make these predictions.

\section{SUMMARY}
\label{sec:conclusion}

Four dwarf irregulars from the \lt\ sample of dwarf galaxies were observed with the \herschel\ PACS and SPIRE photometers to determine the dust properties of typical dwarf galaxies: those with very low metallicity and moderate star formation.  After carefully measuring the flux densities in five FIR bands for each, the emission was compared between the different bands, and compared with previous MIPS 160\um\ data.  Dust masses, temperatures, and emissivities were estimated in these systems from modified blackbody fits to the FIR emission.  This yielded the following results:

\begin{itemize}
\item The fitted dust temperatures of the \lt\ galaxies show a mix of cool and warm dust components.  Previous work on the DGS dwarfs showed that many of the lowest-metallicity galaxies had dust temperatures between $\sim$20--50 K. These new data indicate that very low metallicity dwarfs with moderate star formation have temperatures typically ranging between 18--26 K.

\item All of the these galaxies have extremely low $L_{\rm TIR}$ and dust masses, with $M_d$ ranging from $\sim 10^4 M_\odot$ down to a few hundred $M_\odot$ -- consistent with the lower range of $M_\mathrm{d}$ values measured for galaxies in other surveys.  
The dust masses follow the same trend of decreasing with lower metallicity as seen in the KINGFISH and DGS samples.

\item The integrated $M_\mathrm{d}$ and $L_{\rm TIR}$ of the galaxies studied here are dominated by one or two bright localized regions.  Resolving the emission of other faraway dwarf galaxies is important to avoid misinterpretation of their bulk properties. 

\item We show that the the dust temperature does not follow a trend with $Z$ -- in agreement with the DGS results.

\item Dust-to-gas ratios in the \lt\ sample are smaller than for more massive and metal-rich galaxies and are mostly consistent with the metallicity trends noted for other low-$Z$ dwarfs with low star formation rates. The inferred DGR for DDO 69 is one of the lowest seen to date, at  $1.3 \times 10^{-6}$.  Even ignoring molecular gas, the dust-to-\hi\ ratios of these galaxies generally fall below the linear DGR trend seen in higher-metallicity systems.

\item Comparison with theoretical models suggests a range of different star formation histories could explain the differences in DGR--$Z$ for most of the \lt\ dwarfs studied in this work.  Changing the assumptions of dust sources and sinks (destruction rates) can also predict the DGR for these galaxies without needing bursty SFHs which appear more appropriate for the more highly star-forming (on average) DGS sample. To reach the lowest DGR values observed in this sample, a model galaxy with delayed star formation history, low initial gas mass, extremely efficient dust destruction and reduced dust contributions from stars and interstellar grain growth is needed, though they could be consistent with some less extreme models given the uncertainties.  These dwarf galaxies are likely forming very little dust via stars or grain growth, and have very high destruction rates.  

\end{itemize}

Although this is a relatively small sample of four sources, these new data have helped to populate the sparsely filled parameter space of extremely metal-poor, low surface brightness galaxies.  Our finding that a small number of localized regions of the galaxy contribute a significant amount of the infrared emission and dust mass is relevant for observations of more distant galaxies, highlighting the importance of resolving the desired features in these systems.

\acknowledgments{Funding for this project was provided by NASA JPL RSA grant 1433776.  PJC and HLG acknowledge support from the European Research Council (ERC) in the form of Consolidator Grant {\sc CosmicDust} (ERC-2014-CoG-647939, PI H\,L\,Gomez). SCM acknowledges support from the Agence Nationale de la Recherche (ANR) through the programme SYMPATICO (Program Blanc Projet ANR-11-BS56-0023), and from the EU FP7 project DustPedia (Grant No. 606847). This work was supported by the Programme National ``Physique et Chimie du Milieu Interstellaire'' (PCMI) of CNRS/INSU with INC/INP cofunded by CEA and CNES.  DAH appreciates assistance for publication provided by the National Science Foundation grant AST-1907492.  The National Radio Astronomy Observatory is a facility of the National Science Foundation operated under cooperative agreement by Associated Universities, Inc.}


\facilities{\textit{Herschel Space Telescope}, Very Large Array}

\software{
  \texttt{astropy} \citep{Astropy2018},
  \texttt{ipython} \citep{ipython}, 
  \texttt{lmfit} \citep{lmfit},
  \texttt{matplotlib} \citep{matplotlib},
  \texttt{numpy}  \citep{numpy}, 
  \texttt{SAOImage DS9} \citep{DS9},
  \texttt{Scanamorphos} \citep{Scanamorphos},
  \texttt{scipy} \citep{scipy}.
}

\vspace{2.0cm}

\bibliographystyle{aj}


\end{document}